\documentclass{article}
\usepackage[utf8]{inputenc}
\usepackage[margin=1in]{geometry}
\usepackage[square,numbers]{natbib}
\usepackage{authblk}
\usepackage{bbm}
\usepackage{float}
\usepackage{array}
\usepackage{amsfonts}
\usepackage{amsmath}
\usepackage{amssymb}
\usepackage{amsthm}
\usepackage{mathtools}
\usepackage{color}
\usepackage{appendix}
\usepackage{algorithm}
\usepackage{algpseudocode}
\usepackage[
    colorlinks=true,
    citecolor=blue,
]{hyperref}
\usepackage{enumitem}
\usepackage{cleveref}
\usepackage[font=small,labelfont=bf]{caption}
\usepackage{subcaption}
\usepackage{authblk}

\usepackage{todonotes}

\newcommand{\sD}{\mathcal{D}}
\newcommand{\sF}{\mathcal{F}}

\newcommand{\sH}{\mathcal{H}}
\newcommand{\sI}{\mathcal{I}}
\newcommand{\indic}{\mathbbm{1}}
\newcommand{\br}[1]{\left[#1\right]}
\newcommand{\sbr}[1]{\left\{#1\right\}}
\newcommand{\pr}[1]{\left(#1\right)}
\newcommand{\prob}{\mathbb{P}}
\newcommand{\R}{\mathbb{R}}
\newcommand{\set}[1]{\{#1\}}
\newcommand{\EEE}{\mathbb{E}}
\newcommand{\floor}[1]{\lfloor #1 \rfloor}
\newcommand{\kl}{\mathrm{KL}}
\DeclareMathOperator{\var}{Var}
\DeclareMathOperator{\cov}{Cov}

\DeclareMathOperator{\Binom}{Binom}
\DeclareMathOperator{\Bern}{Bern}
\DeclareMathOperator{\Beta}{Beta}
\DeclareMathOperator{\Exp}{Exp}
\DeclareMathOperator{\Normal}{\mathcal{N}}
\DeclareMathOperator{\subjto}{subject\;to}
\DeclareMathOperator*{\minimize}{minimize\;}

\DeclareMathOperator*{\logit}{logit}
\newcommand{\abs}[1]{\left|#1\right|}
\newcommand{\norm}[1]{\left\lVert#1\right\rVert}

\newtheorem{theorem}{Theorem}
\newtheorem{definition}{Definition}
\newtheorem{lemma}[theorem]{Lemma}

\newtheorem{remark}{Remark}
\newtheorem{example}{Example}

\definecolor{dark-green}{HTML}{228B22}

\title{Guarantees for Comprehensive Simulation Assessment of Statistical Methods}
\author[1,2]{James Yang} 
\author[1]{T. Ben Thompson}
\author[1]{Michael Sklar}
\affil[1]{Confirm Solutions Inc., Stanford, CA, U.S.A.}
\affil[2]{Department of Statistics, Stanford University, Stanford, CA, U.S.A.}

\date{\today}
 
\begin{document}

\maketitle

\begin{abstract}

Simulation can evaluate a statistical method for properties such as Type I Error, FDR, or bias on a grid of hypothesized parameter values. But what about the gaps between the grid-points? Continuous Simulation Extension (CSE) is a proof-by-simulation framework which can supplement simulations with (1) confidence bands valid over regions of parameter space or (2) calibration of rejection thresholds to provide rigorous proof of strong Type I Error control. 
CSE extends simulation estimates at grid-points into bounds over nearby space using a model shift bound related to the Renyi divergence, which we analyze for models in exponential family or canonical GLM form. CSE can work with adaptive sampling, nuisance parameters, administrative censoring, multiple arms, multiple testing, Bayesian randomization, Bayesian decision-making, and inference algorithms of arbitrary complexity. As a case study, we calibrate for strong Type I Error control a Phase II/III Bayesian selection design with 4 unknown statistical parameters. Potential applications include calibration of new statistical procedures or streamlining regulatory review of adaptive trial designs. Our open-source software implementation \textit{imprint} is available at \href{https://github.com/Confirm-Solutions/imprint}{https://github.com/Confirm-Solutions/imprint}.
\end{abstract}


\section{Introduction}

Modern experiment designs can reduce the cost or length of clinical trials, with approaches such as adaptive stopping, multiple testing, and Bayesian decision-making. But, regulated experiments require design properties such as Type I Error and bias to be well-established. 
Analyzing the properties of an ambitious design may require specialized methods and software, and regulators have historically regarded simulations as a lesser form of evidence of design quality.

This paper will demonstrate how  comprehensive guarantees for a wide class of statistical procedures can be built on a simulation backbone. We provide an open-source implementation available at \href{https://github.com/Confirm-Solutions/imprint}{https://github.com/Confirm-Solutions/imprint}, which can provide objective verification of the properties of compatible designs. The techniques we develop can also offer finite-sample conservative confidence intervals and performance analysis for methods including logistic modeling and parameterized survival analysis.

To introduce our setting and terminology, it may be helpful to discuss an example. In Section \ref{ssec:lewis} 
we will use our method, Continuous Simulation Extension (CSE), to calibrate a complex Bayesian design. The outcome of each patient in arm $i$ is binary, modeled as independent $Bernoulli(p_i)$, and the space of unknown statistical parameters is therefore $(p_0, p_1, p_2, p_3) \in [0,1]^4$.
This Bayesian design would be very difficult to analyze mathematically, as it has interim decisions for stopping or adaptively dropping arms and repeatedly uses Bayesian posterior calculations within a hierarchical model. Nevertheless we shall seek to analyze the Type I Error behavior of this design across the null hypothesis space, and calibrate the decision rule to achieve an overall guarantee of Type I Error control. Using CSE's Calibration procedure, we run large-scale simulations to ensure strong Type I Error control over the 4-dimensional cube $(p_0, p_1, p_2, p_3) \in [27\%,76\%]^4$
(see~\Cref{ssec:lewis,sec:calibration} and Appendix \ref{sec:appendix-d} for full details). 
The guarantee level is set to 2.5\%. We give evidence in Section \ref{ssec:lewis} that the calibration process likely results in a Type I Error 
of about 2.3\%, or better if we had run a larger simulation - but the overall 2.5\% guarantee accounts for uncertainty in the calibration process itself and is valid-in-finite-samples over the full region $[23\%,76\%]^4$. 


Section \ref{sec:regulatory-background} discusses related literature and regulatory use of simulations for Type I Error control of clinical trials.

Section \ref{sec:introduction} describes CSE and its two procedure options, Validation and Calibration.

Section \ref{sec:cse}  analyzes the Tilt-Bound and compare against alternative inequalities.

Section \ref{sec:cse-application-workflow} discusses further setup and terminology for applying CSE Validation and Calibration.

Sections \ref{sec:validation} and \ref{sec:calibration} detail the Validation and Calibration procedures, respectively.

Section \ref{sec:confidence-set} discusses how Calibration can offer conservative simulation-based confidence intervals.

Section \ref{sec:cse-for-adaptive-design} discusses handling adaptive designs with CSE.

Section \ref{sec:application-examples} performs CSE for two case studies. In \ref{ssec:berry} we use Validation to analyze a Bayesian basket trial design similar to Berry et al (2013) \cite{berry2013bayesian}. In \ref{ssec:lewis} we Calibrate for strong Type I Error control a complex Phase II-III four-arm Bayesian design with interim decisions based on a hierarchical model, and possible arm-dropping.

Proofs are deferred to the Appendix.

\section{Background and Regulatory Context on the use of Simulation in Trial Design}\label{sec:regulatory-background}

As part of proposing a statistical procedure, general wisdom (and sometimes regulatory guidance \cite{us1998statistical}) dictate that the Type I Error, power, FDR, and/or bias must be analyzed. But, as stated in an FDA experience review \cite{lin2016cber}, ``for less well understood adaptive approaches, the major design properties (eg, type I error rate and power) often cannot be assessed through analytical derivations." 
When other methods fail, simulation is a robust fallback; at the very least, it offers analysis under specific assumed scenarios.  FDA guidance includes advice on the use of simulations to evaluate Type I Error and other operating characteristics of clinical trial designs \cite{food2019adaptive, food2010bayesian}, and regulatory teams such as the Complex Innovative Design program \cite{FDAbrochure} have validated designs with simulation assessment.

This recent progress comes despite a historical view that simulation could not match the guarantees of mathematical analysis. Grieve 2016
\cite{grieve2016idle}, citing Wang and Bretz 2010 \cite{wang2010adaptive}, summarizes: ``The view that control of the type I error is unprovable by simulation was
re-iterated in a panel discussion on adaptive designs at the Third Annual
FDA/DIA Statistics Forum held in Washington in 2009."
\citet{collignon2018adaptive} discuss ``[the regulatory perspective of the EMA as of 2018] as regards the implementation of adaptive designs in confirmatory clinical trials,” and state: “A special need, justified from the context of the trial, would be required to justify a design where type I error rate control can only be ensured by means of statistical simulations, in particular where a design would be available in which type I error rate control can be ensured analytically.”

Surprisingly, we show that \textit{analytically valid proofs of Type I Error control can be constructed from a backbone of simulation}. In addition, we provide software to offer rigorous affirmative answers to several questions of regulatory interest for simulation assessment, such as:

\begin{itemize}

\item Have enough simulations been completed to establish the performance target?

\item Is the gridding of simulation parameters adequately fine? 

\item Given the possibility of selection of the scenarios by motivated designers, or prior constrained optimization of the design which could move Type I Error to where it is not being checked - are the overall results sufficiently representative?
\end{itemize}

Similar questions were raised in a public FDA workshop in March 2018 \cite{FDAworkshop}. CBER/CDER 2019 draft guidance on adaptive design \cite{food2019adaptive} says of simulations: ``In many cases, it will not be possible to estimate Type I error probability for every set of null assumptions even after taking clinical and mathematical considerations into account. It is common to perform simulations on a grid of plausible values and argue based on the totality of the evidence from the simulations that maximal Type I error probability likely does not exceed a
desired level across the range covered by the grid." But what constitutes such a sufficient totality of evidence is up to interpretation. According to the FDA Complex Innovative Design Program's 2021 progress report \cite{price2021us}, additional simulations or scenarios were requested by regulators in 3 out of the 5 case studies considered.
As Campbell (2013) states, “there is an extra regulatory burden for a Bayesian or adaptive submission at the design stage. It is simply a lot more work to review and to ask the sponsor to perform simulations for additional scenarios and in some cases the reviewer may need to reproduce the simulation results.” \cite{campbell2013similarities} We hope that for applicable trials, CSE software can allow for objective verification of standards and streamline this process.

A common goal in stochastic simulation with unknown input parameters is to integrate the operating characteristics over the uncertainty \cite{barton2022input}, using techniques such as Bayesian analysis \cite{chick2001input,zouaoui2004accounting}, bootstrapping \cite{barton1993uniform,barton2014quantifying}, or the delta method \cite{cheng1998two,cheng2004calculation}. 
However in pivotal medical trials, rather than integrating over the uncertainty of the treatment effect(s), regulators typically request worst-case control of the Type I Error. Justifications include protection against over-optimism in the use of expert judgment or Bayesian priors; regulators' resource and human capital constraints which favor the use of simple and sweeping guarantees; bluff-prevention incentives, because the large ratio between trial costs and profits could incentivize running trials on poor drug candidates unless a minimum frequentist condition is met (\cite{tetenov2016economic, bates2022principal}); and idiosyncrasies of new treatments which can lead to challenges in defining appropriate prior distributions or uncertainty widths as inputs to such an integration. This is not to say that Bayesian methods are ruled out or discouraged - see \cite{food2010bayesian,EMA-QandA,campbell2013similarities}.

Nonparametric estimation tools including approaches such as kriging and Gaussian  Processes are commonly seen across many disciplines for analyzing simulation outputs \cite{kleijnen2009kriging,iskandarani2016overview,schefzik2013uncertainty}. In clinical trial simulations, the `adaptr' package uses Gaussian Process modeling of the Type I Error surface to calibrate the Type I Error of adaptive and Bayesian designs.\cite{granholm2022adaptr} These nonparametric approaches often 
build a probabilistic meta-model for the response surface $f$ and can provide effective interpolations. Yet because $f$ typically has a fixed ground truth the modeling assumptions for interval coverage may be false or unverifiable.


For the task of determining the worst-case performance, the Distributionally Robust Optimization (DRO) literature contains several works which involve simulation to assess worst-case performance of a model over confidence sets. \cite{hu2012robust} considers a climate model with uncertain multivariate Gaussian parameters $\mu, \Sigma$ and seeks the worst-case performance; \cite{glasserman2014robust} considers metrics such as Value-at-Risk using simulations and model shifts of Gaussian financial models; and \cite{ghosh2019robust} takes a general simulate-and-discretize approach and attempt non-convex optimization, although convergence to a global optimum may be hard to verify. In contrast to these works, our approach yields guarantees in finite samples for general experiments within GLM and canonical exponential family model classes. 

Our bounding approach is particularly well-suited for statistical experiments: 
counter-intuitively, the presence of noise is helpful to ensure the computation is feasible. 
The Renyi divergence quantities required will typically be finite and and support sufficiently-sharp inequalities; rapid or non-smooth ``phase changes" in the target metric will not occur over regions where the target metric depends smoothly on $\theta$, due to the smooth sample likelihood of GLM and exponential families. These favorable conditions may not hold for simulations in other disciplines, such as deterministic physical systems.

Although we introduce new analysis to achieve our task, similar theoretical tools have long existed. For example, Pinsker's inequality can explicitly bound changes in the probabilities of sets at nearby parameter values; but Pinsker's inequality \cite{tsybakov2009introduction} is highly inefficient compared to our Tilt-Bound - see our Section \ref{ssec:inequality-comparison}. Our precursor work \cite{sklar:2022} used a conservative Taylor Expansion bound, although according to our investigation in Section \ref{ssec:inequality-comparison} the Tilt-Bound is again superior.

The Tilt-Bound can be written in terms of Renyi divergences, and is similar to bounds in other works such as \cite{esposito2021generalization} Corollary 3, and \cite{begin2016pac} Theorem 9. \cite{nielsen2011r} computes Renyi divergences for exponential families. \cite{van2014renyi} establishes convexity of the Renyi divergence $D_{\alpha}(P || Q)$ in $P$ and $Q$, a result which applies to mixture distributions but not the parametric families we study. Our Theorem \ref{thm:tilt-bound-qcp} establishes a novel quasi-convexity result for the Tilt-Bound which allows for computationally easy worst-case control over volumes of space.





\section{Introduction to CSE}\label{sec:introduction}

Superseding Chapter 5 of~\citet{sklar:2022}, 
we propose bounds for error quantification of simulation results and develop analysis guarantees covering bounded \textit{regions} of the input parameter space. We use the term
\emph{Continuous Simulation Extension} (CSE) 
to refer to the mathematical tools developed in in~\Cref{sec:cse}
which we use for two concrete procedures, Validation (~\Cref{sec:validation}) and Calibration (~\Cref{sec:calibration}):
\begin{enumerate}[align=left]
\item[\textbf{Validation}:] 
(~\Cref{sec:validation}) Given a fixed design, provide a lower or upper-bound estimates,
$(\hat{\ell}(\cdot), \hat{u}(\cdot))$,
of an operating characteristic (such as Type I Error, false-discovery rate (FDR), or
bias of bounded estimator)
over a (bounded) parameter space
with a pointwise-valid confidence bound with confidence parameter $\delta$. Thus, we have 
\begin{align*}
\forall \theta \in \Theta,\, 
\prob\pr{f(\theta) \leq \hat{u}(\theta)} \geq 
1 - \delta \text{ and }
\prob\pr{f(\theta) \geq \hat{\ell}(\theta)} \geq  1- \delta
\end{align*}
where $\Theta$ is a bounded region of parameter space
and $f(\theta)$ is the unknown operating characteristic of interest.

\item[\textbf{Calibration}:]
(~\Cref{sec:calibration}) Calibrate the critical threshold of a design 
to achieve a provable bound 
on the expected Type I Error of the selected threshold
on a (bounded) parameter space.
That is, we show how to select a (random) critical threshold, 
denoted $\hat{\lambda}^*$, such that 
\begin{align*}
\forall \theta \in \Theta,\, 
\EEE\br{f_{\hat{\lambda}^*}(\theta)} 
\leq 
\alpha
\end{align*}
where $f_{\lambda}(\theta)$ is the Type I Error of the design 
with critical threshold $\lambda$ under parameter $\theta$,
and $\Theta$ is the null hypothesis region of parameter values.

\end{enumerate}

As inputs to these basic tasks, we require (1) a modeling family for outcome generation parametrized by the unknown vector $\theta \in \Theta$, with a likelihood that is exponential family or canonical form GLM with respect to the parameter $\theta$, (2) a protocol plan which specifies the statistical procedure to be performed to analyze the data, (3) repeated simulations of the data and procedure taken from a fine grid of parameter values $\theta_j$ for $j \in 1 , \ldots, J$.


Where CSE is applicable, the guarantees of calibration and validation are finite-sample valid and conservative. However, many simulations may be required to reduce its conservative slack. Increasing the density and number of simulations typically reduces the width of validation bands and improves the average power of calibration tests. 

CSE provides an alternative to mathematical analysis for deriving valid error guarantees and confidence intervals. For example, as we discuss in Section \ref{sec:confidence-set}, due to the equivalence of statistical tests and confidence intervals, CSE calibration can reduce the problem of deriving a conservative upper confidence interval around an estimator $\hat{\theta}(X)$ to merely ``guessing" a data-dependent relative width:

$$\left( -\infty, \hat{\theta}(X) + \lambda u(X) \right)$$ 

where $\lambda$ is a tunable scaling factor which will be calibrated to correct the worst-case error in its coverage probability.


We should clarify that the space $\Theta$ of unknown parameters must consist of \textit{statistical parameters} which are connected to the simulation outcomes through a likelihood model. We contrast statistical parameters to \textit{design parameters} such as maximum sample sizes for the protocol, which we assume are fixed or conditioned on and thus removed as dimensions from the simulation model. Covariates could be viewed as either statistical parameters (if analyzed as part of the random model) or, preferably, as design parameters (if conditioned-on). 


Fortunately, the presence of complications such as interim analyses, highly adaptive stopping, or complex algorithms may affect the number of simulations required only minimally (though a complex algorithm may naturally extend the processing time of each individual simulation). This scaling is favorable for highly adaptive or bandit-like designs, where other ``brute force" methods such as dynamic programming may struggle with the large space of possible states. Intuitively, CSE is ``brute forcing" the space of input parameters, but not the possible states of the simulation.

\section{The Tilt-Bound for CSE}\label{sec:cse}


In this section we introduce a key technical tool for
\emph{Continuous Simulation Extension} (CSE).
The CSE approach, described further in \Cref{sec:cse-application-workflow}, uses simulation to estimate the performance over an input grid of model parameter values, and then bounds the extent to which that performance can vary across nearby space.
To perform bounding, CSE uses an inequality we call 
the \emph{Tilt-Bound}.
Given the family of distributions $P_\theta$ 
for the underlying data,
a ``knowledge point'' $\theta_0$,
and a ``target point'' $\theta$,
the Tilt-Bound gives a deterministic upper bound
on the change of operating characteristics such as probabilities 
of an arbitrary event 
(e.g. Type I Error or power of any design),
false-discovery rate (FDR),
or bias of bounded estimators
assuming knowledge of the outcome metric at $\theta_0$.

In~\Cref{ssec:tilt-bound-properties},
we state the guarantee of the Tilt-Bound and its properties.
In~\Cref{ssec:tilt-bound-normal},
we provide an explicit treatment of the Tilt-Bound
on the Normal location family
$\set{\Normal\pr{\theta, 1} : \theta \in \Theta}$
as a concrete example.
In~\Cref{ssec:inequality-comparison},
we perform a brief numerical comparison of the 
quality of the Tilt-Bound against Pinsker's Inequality 
and the Taylor Expansion approach as in~\citet{sklar:2022} to demonstrate
the tightness of the Tilt-Bound.
In~\Cref{sec:cse-application-workflow},
we outline the general workflow of applying CSE
to prepare for the validation and calibration procedures in~\Cref{sec:validation,sec:calibration}. 

\subsection{Tilt-Bound and its Properties}
\label{ssec:tilt-bound-properties}

We first state the Tilt-Bound and prove its guarantee in~\Cref{thm:tilt-bound}.

\begin{theorem}[Tilt-Bound]\label{thm:tilt-bound}
Let $\set{P_{\theta} : \theta \in \Theta}$
denote a family of distributions with density
\begin{align}
    p_\theta(x) = \exp\br{g_\theta(x) - A(\theta)}
    \label{eq:general-dist-assumption}
\end{align}
for some functions $g_{\theta}(x) := g(\theta,x)$ and $A(\theta)$ such that $\int p_{\theta}(x) d\mu(x) = 1$
for some base measure $\mu$. 
Denote
\begin{align}
    \Delta_{\theta}(v, x)
    &:=
    g_{\theta+v}(x) - g_{\theta}(x)
    \label{eq:tilt-bound:Delta-def}
    \\
    \psi(\theta, v, q)
    &:=
    \log \EEE_{\theta}\br{e^{q\Delta_{\theta}(v, X)}}
    \label{eq:tilt-bound:psi-def}
\end{align}
Suppose $F: \R \to [0,1]$ is a measurable function
and $f(\theta) := \EEE_\theta\br{F(X)}$
where $X \sim P_\theta$.
Fix any ``knowledge point" $\theta_0 \in \Theta$ 
and ``target point" $\theta_0 + v \in \Theta$. 
Then, for any $q \in [1,\infty]$,
\begin{align}
    f(\theta_0 + v)
    &\leq
    f(\theta_0)^{1-\frac{1}{q}}
    \exp\br{
    \frac{\psi(\theta_0, v, q)}{q}
    - \psi(\theta_0, v, 1) 
    }
    \label{eq:tilt-boundddd}
\end{align}
The bound~\labelcref{eq:tilt-boundddd}
is called the \emph{Tilt-Bound}, denoted as:
\begin{align}
    U(\theta_0, v, q, a) 
    := 
    a^{1-\frac{1}{q}}
    \exp\br{
        \frac{\psi(\theta_0, v, q)}{q}
        - \psi(\theta_0, v, 1)
    }
    \label{eq:tilt-bound:def}
\end{align}

\begin{proof}
Fix any $1 \leq q \leq \infty$ and define its H\"{o}lder conjugate $p := \pr{1-q^{-1}}^{-1}$, with the convention that 
$p=\infty$ if $q = 1$ and vice versa.
Define
\[
    \ell_\theta(x)
    :=
    \log p_\theta(x) 
    =
    g_\theta(x) - A(\theta)
\]
Fix any $\theta_0 \in \Theta$.
Then, for any $\theta \in \Theta$,
\begin{align*}
    f(\theta)
    &=
    \EEE_{\theta}\br{F(X)}
    =
    \EEE_{\theta_0}\br{
        F(X)e^{\ell_{\theta}(X)-\ell_{\theta_0}(X)}
    }
    \\&\leq
    \norm{F(X)}_{L^p(P_{\theta_0})}
    \norm{e^{\ell_{\theta}(X)-\ell_{\theta_0}(X)}}_{L^q(P_{\theta_0})}
    \\&=
    \norm{F(X)}_{L^p(P_{\theta_0})}
    \norm{e^{g_{\theta}(X)-g_{\theta_0}(X)}}_{L^q(P_{\theta_0})}
    e^{-\pr{A(\theta) - A(\theta_0)}}
    \\&\leq
    f(\theta_0)^{\frac{1}{p}}
    \pr{
    \EEE_{\theta_0}\br{
        e^{q\pr{g_{\theta}(X) - g_{\theta_0}(X)}}
    }}^{\frac{1}{q}}
    e^{-\pr{A(\theta) - A(\theta_0)}}
\end{align*}
Noting that
\[
\psi(\theta_0, v, 1) \equiv 
A(\theta_0+v) - A(\theta_0)
\]
we have that for any $v\in \Theta - \theta_0$,
\begin{align*}
    f(\theta_0 + v)
    &\leq
    f(\theta_0)^{1-\frac{1}{q}}
    \exp\br{
        \frac{\psi(\theta_0, v, q)}{q} 
        - \psi(\theta_0, v, 1)
    }
\end{align*}
\end{proof}
\end{theorem}

Our primary use-case of~\Cref{thm:tilt-bound}
is when $F(x)$ is the indicator that a design $\sD$ rejects
the null using data $x$, so that $f(\theta)$ is the Type I Error
whenever $\theta$ is in the null-space.
\Cref{thm:tilt-bound} then says that given the knowledge 
of the Type I Error at a parameter $\theta_0$
and the family of distributions $\set{P_\theta : \theta \in \Theta}$,
we can provide a (deterministic) upper bound 
of the Type I Error at a new point $\theta$.

\begin{remark}[Extension to Bounded Functions]
\label{rem:extension-bounded}
To apply the Tilt-Bound in the case of a function $F$ bounded in $[a,b]$, we may simply standardize the range to $[0,1]$ with a shift and scale and apply~\Cref{thm:tilt-bound}
to the rescaled target 
$\tilde{F}(x) := \frac{F(x)-a}{b-a}$ to get
\begin{align*}
    \frac{f(\theta_0+v) - a}{b-a}
    &\leq
    \pr{
    \frac{f(\theta_0) - a}{b-a}
    }^{1-\frac{1}{q}}
    \exp\br{
        \frac{\psi(\theta_0, v, q)}{q}
        - \psi(\theta_0, v, 1)
    }
\end{align*}
Rearranging,
\begin{align*}
    f(\theta_0+v)
    \leq
    a + (b-a) 
    \pr{
    \frac{f(\theta_0) - a}{b-a}
    }^{1-\frac{1}{q}}
    \exp\br{
        \frac{\psi(\theta_0, v, q)}{q}
        - \psi(\theta_0, v, 1)
    }
\end{align*}
\end{remark}

\begin{remark}[Extension to Lower Bound]
\label{rem:extension-lower}
Although the Tilt-Bound is stated as an upper bound, 
one can easily extend the result of~\Cref{thm:tilt-bound} to construct a lower bound.
Indeed, given $F : \R \to [0,1]$,
we may consider $\tilde{F} := 1 - F$ 
and apply~\Cref{thm:tilt-bound} to get that
\begin{align*}
    1 - f(\theta_0 + v)
    &\leq
    \pr{1-f(\theta_0)}^{1-\frac{1}{q}}
    \exp\br{
        \frac{\psi(\theta_0, v, q)}{q}
        - \psi(\theta_0, v, 1)
    }
\end{align*}
Rearranging,
\begin{align*}
    f(\theta_0+v)
    \geq
    1 - \pr{1-f(\theta_0)}^{1 - \frac{1}{q}}
    \exp\br{
        \frac{\psi(\theta_0, v, q)}{q} 
        - \psi(\theta_0, v, 1)
    }
\end{align*}
\end{remark}

\begin{remark}[Optimized Tilt-Bound]\label{rem:optimized-tilt-bound}
The Tilt-Bound is based on a direct application of H\"{o}lder's Inequality to develop a \emph{family of bounds} 
indexed by a parameter $q \in [1, \infty]$.
Since the bound holds for every $q \in [1,\infty]$,
it can also be minimized over $q$ for sharper results. 
That is, we have that
\begin{align}
    f(\theta_0 + v)
    \leq 
    \inf\limits_{q \in [1,\infty]}
    U(\theta_0, v, q, f(\theta_0))
    \label{eq:optimized-tilt-bound}
\end{align}
Similarly, we also have the optimal worst-case bound 
over a space $H \subseteq \Theta - \theta_0$
\begin{align}
    \sup\limits_{v \in H} 
    f(\theta_0 + v)
    \leq 
    \inf\limits_{q \in [1,\infty]}
    \sup\limits_{v \in H}
    U(\theta_0, v, q, f(\theta_0))
    \label{eq:optimized-tilt-bound-tile}
\end{align}
We refer to results~\labelcref{eq:optimized-tilt-bound,eq:optimized-tilt-bound-tile} as
the \emph{Optimized Tilt-Bound}
and the \emph{Tilewise Optimized Tilt-Bound}, respectively.
These optimized Tilt-Bounds
will be useful in~\Cref{sec:validation,sec:calibration}.
\end{remark}

Both ~\labelcref{eq:optimized-tilt-bound} and ~\labelcref{eq:optimized-tilt-bound-tile} are optimization problems; ~\Cref{thm:tilt-bound-qcp,thm:tilt-bound-qcv} show
these will be straightforward to evaluate.
Indeed,~\Cref{thm:tilt-bound-qcp} shows that the minimization problem in both~\labelcref{eq:optimized-tilt-bound,eq:optimized-tilt-bound-tile} are one-dimensional
quasi-convex minimization problems in general.
With an additional ``linearity'' condition 
of the family of distributions,~\Cref{thm:tilt-bound-qcv}
shows that the Tilt-Bound is quasi-convex
as a function of the displacement $v$ as well.
Therefore if $H$ is a polytope,
the supremum in~\labelcref{eq:optimized-tilt-bound-tile} can be reduced to a (finite) maximum
of the Tilt-Bound on the vertices of $H$.
We state~\Cref{thm:tilt-bound-qcv,thm:tilt-bound-qcp} and leave the proof in~\Cref{appendix:thm:tilt-bound-qcv,appendix:thm:tilt-bound-qcp}.

\begin{theorem}[Quasi-convexity in $q$ of the Tilt-Bound]%
\label{thm:tilt-bound-qcp}
Let $\set{P_{\theta} : \theta \in \Theta}$
be as in~\labelcref{eq:general-dist-assumption} and
$U$ be as in~\labelcref{eq:tilt-bound:def}.
Fix any $\theta_0 \in \Theta \subseteq \R^d$,
a set $S \subseteq \R^d$,
and $a \geq 0$.
Assume that for all $v \in S$,
$\Delta_{\theta_0}(v, X)$ in~\labelcref{eq:tilt-bound:Delta-def}
is not constant $P_{\theta_0}$-a.s..
Then, $W(q) := \sup\limits_{v \in S} U(\theta_0, v, q, a)$ is quasi-convex and 
there exists a global minimizer 
$q^\star \equiv q^\star(\theta_0, v, a) \in [1, \infty]$.
Moreover, if $a > 0$, $S$ is finite,
and $W$ is not identically infinite,
then $W(q)$ is strictly quasi-convex 
and the minimizer $q^\star$ is unique.
\end{theorem}

\begin{theorem}[Quasi-convexity in $v$ of the Tilt-Bound]%
\label{thm:tilt-bound-qcv}
Consider the setting of~\Cref{thm:tilt-bound}.
Fix any $\theta_0 \in \Theta$.
Suppose that $\Delta(v, x) \equiv \Delta_{\theta_0}(v, x)$ 
in~\labelcref{eq:tilt-bound:Delta-def}
is linear in $v$, i.e.
$\Delta(v, x) = W(x)^\top v$ 
for some vector $W(x) \in \R^d$.
Then, the Tilt-Bound~\labelcref{eq:tilt-bound:def} 
is quasi-convex as a function of $v$.
\end{theorem}

Note that~\Cref{thm:tilt-bound} holds 
generally with minimal assumptions. 
In particular,~\Cref{thm:tilt-bound} 
applies to two large classes of models: 
exponential families, and canonical generalized linear models (when conditioning on covariates).
We discuss these in ~\Cref{ex:exp-fam,ex:glm}.
It is worth emphasizing that~\Cref{thm:tilt-bound,thm:tilt-bound-qcp,thm:tilt-bound-qcv}
can apply to truly arbitrary tests.
That is, rejections may be based on any 
parametric, non-parametric, Bayesian, or black-box methods. The Tilt-Bound will be valid in any case; in practice, the key question will be whether the parametric data generation model is sufficiently well-specified that the guarantee is relevant.


\begin{example}[Exponential Family]\label{ex:exp-fam}
Suppose $\set{P_\theta : \theta \in \Theta}$
is an exponential family with natural parameter $\theta$
so that the density is of the form
\begin{align*}
    p_{\theta}(x)
    &=
    \exp\br{g_\theta(x) - A(\theta)}
\end{align*}
where $g_\theta(x) = T(x)^\top \theta$.
Then, the conditions for~\Cref{thm:tilt-bound} hold.

Further, the Tilt-Bound simplifies to a simple formula.
Indeed, for any $\theta_0 \in \Theta$,
\begin{align*}
    \Delta(v, x) 
    = g_{\theta_0+v}(x) - g_{\theta_0}(x)
    = T(x)^\top v
\end{align*}
Since,
\begin{align*}
    \psi(\theta, v, q)
    =
    \log \EEE_\theta\br{e^{q \Delta(v, X)}}
    =
    \log \EEE_\theta\br{e^{q T(X)^\top v}}
    =
    A(\theta + qv) - A(\theta)
\end{align*}
the Tilt-Bound simplifies to
\begin{align}
    U(\theta_0, v, q, a) 
    &=
    a^{1-\frac{1}{q}}
    \exp\br{
        \frac{A(\theta_0+qv)-A(\theta_0)}{q}
        - \pr{A(\theta_0+v) - A(\theta_0)}
    }
    \label{eq:tilt-bound:exp-fam}
\end{align}
Since $\Delta(v, X)$ is not constant 
$P_{\theta_0}$-a.s. for any $v$,
we may apply~\Cref{thm:tilt-bound-qcp}.
Moreover, since $\Delta$ is linear in $v$, 
we are in position to apply~\Cref{thm:tilt-bound-qcv} as well.

In~\labelcref{eq:tilt-bound:exp-fam}, we see that the exponent term
is a difference of directional secants.
As difference of slopes contains Hessian information,
it is intuitive to think about the exponent term
as ``curvature information'' in $P_\theta$.
In fact, under the (natural parameter) 
exponential family,
the Hessian of $A(\theta)$ is precisely the 
negative of the Fisher's Information.
The name ``Tilt-Bound'' comes from the 
fact that the exponent term mimics exponential-tilts
in the theory of exponential family~\citep{butler:2007,siegmund:1976}.
\end{example}

\begin{example}[Canonical Generalized Linear Model (GLM)]
\label{ex:glm}
Under the canonial GLM framework,
we model each response $y_i$ as exponential
family where the natural parameter 
is parameterized as $x_i^\top \theta$
(fixed $x_i$'s).
Hence, letting $y \in \R^n$ be the vector of responses
and $X \in \R^{n \times d}$ denote the matrix of 
covariates with each row as $x_i^\top$,
the density of $y$ is of the form
\begin{align*}
    p_{\theta}(y)
    &=
    \exp\br{
        g_\theta(y)
        - A(\theta)
    }
\end{align*}
with $g_\theta(y) := y^\top X \theta$.
Then, the conditions for~\Cref{thm:tilt-bound} hold.

Further, for every fixed $\theta_0 \in \Theta$,
\begin{align*}
    \Delta(v, y)
    &=
    g_{\theta_0 + v}(y)
    -
    g_{\theta_0}(y)
    =
    y^\top X v
    = 
    T_X(y)^\top v
\end{align*}
where $T_X(y) := X^\top y$.
Similar to~\Cref{ex:exp-fam}, 
\begin{align*}
    \psi(\theta, v, q)
    =
    \log \EEE_\theta\br{e^{q \Delta(v, X)}}
    =
    A(\theta + qv) - A(\theta)
\end{align*}
so, the Tilt-Bound simplifies to the same form as in~\labelcref{eq:tilt-bound:exp-fam}.
By the same arguments as in~\Cref{ex:exp-fam},
we are in position to apply~\Cref{thm:tilt-bound-qcp,thm:tilt-bound-qcv}.
\end{example}

The Tilt-Bound is closely connected to the R\'{e}nyi divergence, which is defined by
\begin{align*}
    D_\alpha(Q \| P)
    =
    \frac{1}{\alpha - 1}
    \log \EEE_P\br{\pr{\frac{dQ}{dP}}^\alpha}
\end{align*}
for any $\alpha \in (0,1) \cup (1,\infty)$~\citep{van_Erven:2014}.
The R\'{e}nyi divergence can be extended to $\alpha \in \set{0,1,\infty}$ by taking limits.
In the same spirit of Cressie-Read's Power Divergence family of statistics that embeds
many important goodness-of-fit statistics~\citep{cressie:1984},
R\'{e}nyi divergence embeds many important divergences such as the Kullback-Leibler (KL) divergence~\citep{van_Erven:2014,esposito:2021,begin:2016}.
\citet{van_Erven:2014} shows in Theorem 8 that the R\'{e}nyi divergence satisfies a change-of-measure inequality that is precisely
the claim in~\Cref{thm:tilt-bound}.
Indeed, there is an explicit relationship
between the Tilt-Bound and R\'{e}nyi divergence:
\begin{align*}
    U(\theta_0, v, q, a)
    &=
    a^{1-\frac{1}{q}}
    \exp\sbr{
    \pr{1 - \frac{1}{q}}
    D_q(P_{\theta_0+v} \| P_{\theta_0})
    }
\end{align*}
However, to the best of our knowledge, we are not aware of previous results using quasi-convex analysis of the R\'{e}nyi divergence as we do to extend tractable statistical guarantees (such as for Type I Error control).


\subsection{Tilt-Bound on Normal Location Family}
\label{ssec:tilt-bound-normal}

\begin{figure}[t]
    \centering
    \includegraphics[width=0.8\textwidth]{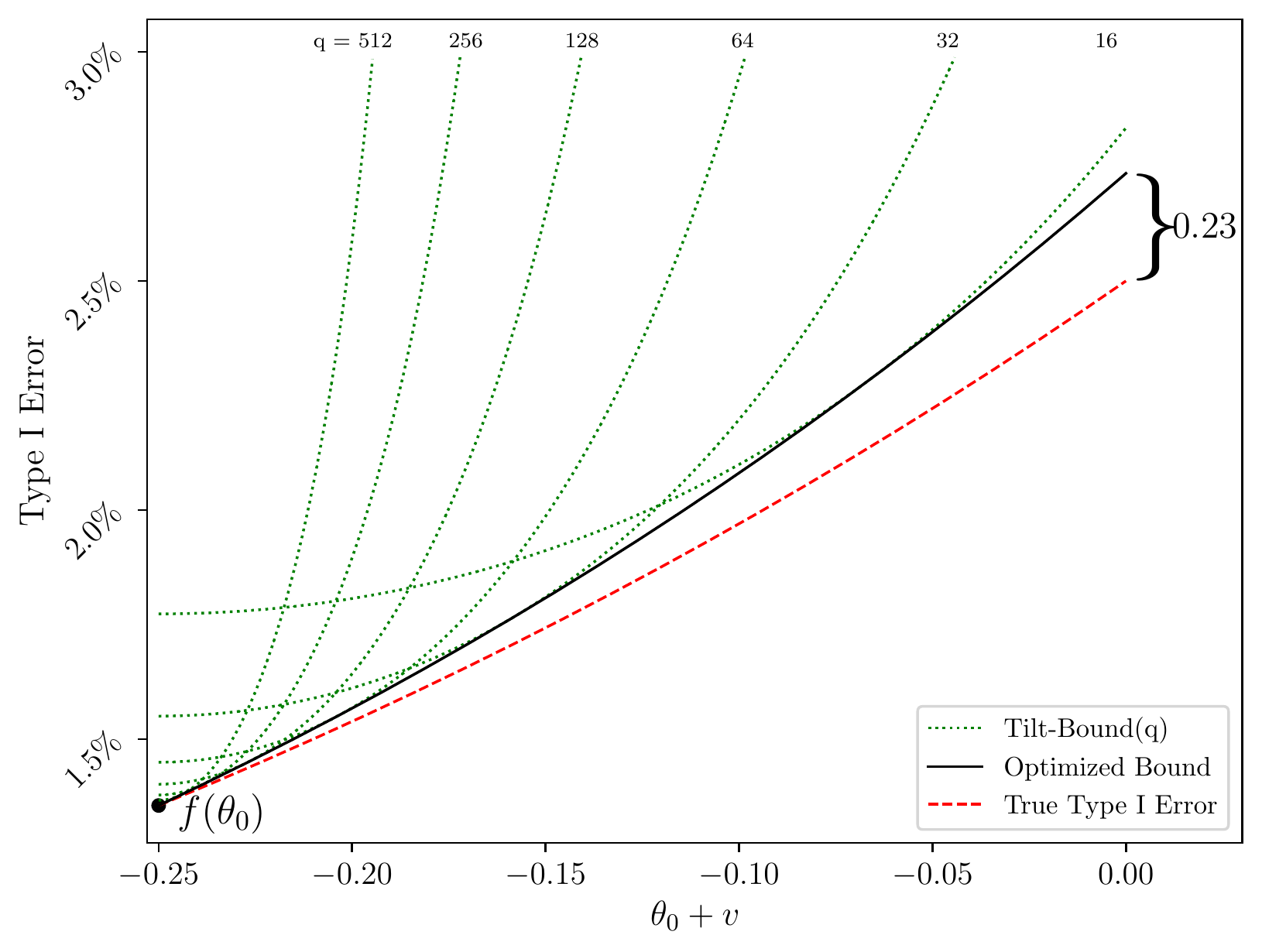}
    \caption{The normal Tilt-Bound~\labelcref{eq:tilt-bound-normal}
    for various fixed 
    values of $q$ (\textbf{\textcolor{dark-green}{green}} dotted lines) 
    are overlaid with the Optimal Tilt-Bound 
    (\textbf{black} solid line),
    which optimizes for $q$ at every $\theta$. 
    It has been assumed that the true Type I Error 
    at the knowledge point $\theta_0 = -0.25$ is known.
    As a baseline comparison, we plot the true Type I Error
    (\textbf{\textcolor{red}{red}} dotted line). 
    In this case, the bound is only mildly conservative, although the slack
    grows with distance from the knowledge point. 
    }
    \label{fig:greens-up}
\end{figure}

Since our primary demonstrative example,
the one-sided z-test,
assumes data under a Normal location family,
we give an explicit treatment of the Tilt-Bound in this setting.
The Normal location family is given by
\begin{align}
\set{\Normal\pr{\theta, 1} : \theta \in \Theta}
\label{eq:normal-family}
\end{align}
which is an exponential family with 
the log-partition function 
$A(\theta) = \frac{\theta^2}{2}$.
Unless stated otherwise,
the demonstrative example is the one-sided z-test for testing 
\[H_0: \theta \leq 0 \quad H_1: \theta > 0\]
where $X \sim \Normal(\theta, 1)$.
Recall that the one-sided z-test rejects if 
$X > z_{1-\alpha}$, the $1-\alpha$ quantile of the 
standard normal distribution.

For the Normal location family~\labelcref{eq:normal-family},
the Tilt-Bound simplifies to:
\begin{align}
    U(\theta_0, v, q, a)
    &=
    a^{1 - \frac{1}{q}} 
    \exp\pr{\frac{(q - 1) v^2}{2}}
    \label{eq:tilt-bound-normal}
\end{align}
In~\Cref{fig:greens-up}, 
we plot the Tilt-Bound for various fixed values of 
$q \in \set{2^4, 2^5, \ldots, 2^9}$ 
and overlay the Optimized Tilt-Bound as discussed in~\Cref{rem:optimized-tilt-bound}.
Here, we assume that the knowledge point
is $\theta_0 := -0.25$
and the Type I Error of the z-test with $\alpha = 0.025$, $f(\theta_0)$, is given.
We see that for any fixed value of $q$ there is some displacement $v$ for which the Tilt-Bound is undesirably loose, but the Optimized Tilt Bound is tight over a wide range of $v$.
Indeed, the largest difference occurs at the furthest target point at $\theta_0 + v = 0$
with a gap of only $0.23\%$.

\subsection{Inequality Comparison}\label{ssec:inequality-comparison}

We demonstrate the quality of the Tilt-Bound approximation
by comparing with existing tools, namely Pinsker's Inequality
and the Taylor approximation given in~\citet{sklar:2022}.

\begin{figure}[t]
    \centering
    \includegraphics[width=0.8\textwidth]{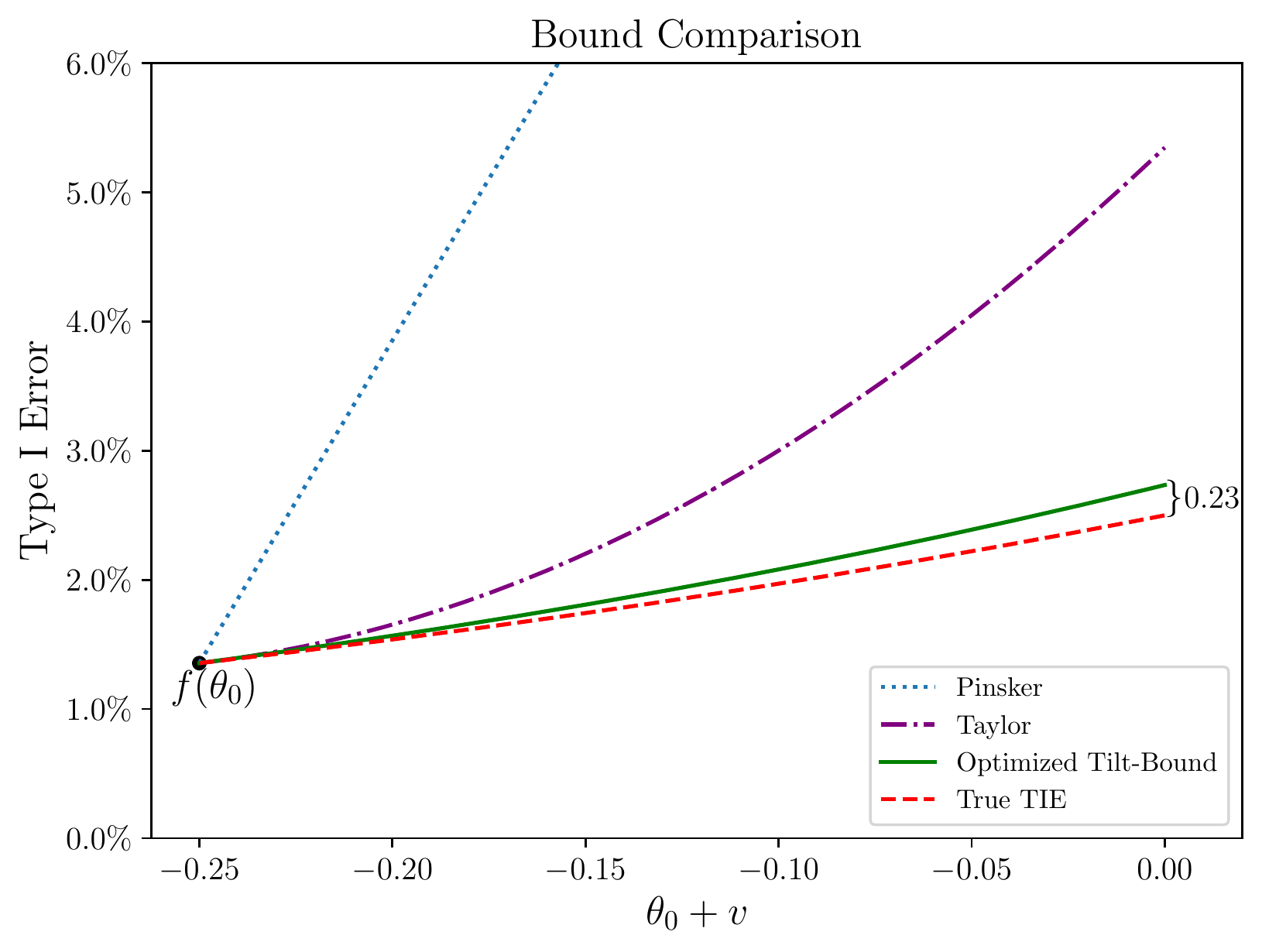}
    \caption{Comparison of methods for constructing an upper bound for the Type I Error. The true (unknown) Type I Error function is the z-test power function,
    $f(\theta) = \Phi\pr{\theta - \Phi^{-1}(1-2.5\%)}$. To create these upper bounds from data at the origin point $\theta_0 = -0.25$, we assume exact knowledge of $f(\theta_0)$. The Taylor Bound also uses exact knowledge of $f'(\theta_0)$ and the upper bound $f''(\theta) \leq 1$ as in \cite{sklar:2022}. 
    }
    \label{fig:bound-comparison}
\end{figure}

Consider the one-sided z-test 
as described in~\Cref{ssec:tilt-bound-normal}
with $\theta_0 = -0.25$ and $\alpha = 0.025$.
In~\Cref{fig:bound-comparison}, we plot the following curves:
\begin{itemize}
\item True Type I Error: 
    \begin{align*}
        f(\theta) = \Phi\pr{\theta - \Phi^{-1}(1-\alpha)}
    \end{align*}
    where $\Phi$ is the standard normal CDF.
    
\item The Optimized Tilt-Bound for the Normal location family~\labelcref{eq:tilt-bound-normal}:
    \begin{align*}
        f(\theta_0 + v) 
        \leq
        \inf\limits_{q \geq 1}
        U(\theta_0, v, q, f(\theta_0)) 
        =    
        \inf\limits_{q \geq 1} 
        \sbr{
            f(\theta_0)^{1 - \frac{1}{q}} 
            \exp\pr{\frac{(q - 1) v^2}{2}}
        }
    \end{align*}

\item The Taylor Expansion Bound used by~\citet{sklar:2022} assuming perfect estimation of the true value $f'(\theta_0)$ and the upper bound $f''(\theta) \leq 1$, which yields
\begin{align*}
    f(\theta_0 + v) \leq f(\theta_0) + f'(\theta_0) v + \frac{v^2}{2}
\end{align*}
\item Pinsker's Inequality:
\begin{align*}
    f(\theta_0 + v) 
    &\leq
    f(\theta_0)
    +
    \sqrt{\frac{1}{2}D_{\kl}(P_{\theta_0} \| P_{\theta_0+v})}
    =
    f(\theta_0) + \frac{\abs{v}}{2}
\end{align*}
\end{itemize}
We observe that the Optimized Tilt-Bound is superior by 
a wide margin and stays close to the true Type I Error. 
At this distance of $0.25$ standard deviation, 
the Optimized Tilt-Bound yields $2.73\%$ where the true Type I Error is $2.5\%$. 
The relative advantage of the Tilt-Bound is consistent
across other choices of origin point $\theta_0$ 
and distance $v$, 
except that the Taylor bound can be better for very small distances. 
However, in practice, the Taylor bound as described in~\citet{sklar:2022} requires conservative estimation of $f'(\theta_0)$ in addition to $f(\theta_0)$, 
which can erase this advantage 
in practical numbers of simulations. 
Thus, we generally recommend the Optimized Tilt-Bound.

We further remark that the bound performance in~\Cref{fig:bound-comparison} will hold for model classes where the likelihood model is asymptotically Gaussian and the test of interest is asymptotically equivalent to a z-test. 
Indeed, an approximate z-test can then be recovered under an appropriate re-mapping of the parameters 
to the signal-to-noise scale, in which case, 
these inequalities are also recoverable 
with an appropriate scaling
(assuming the likelihoods and bounds also converge to those of Gaussians). 
Even more generally, in a multi-dimensional asymptotically Gaussian parametric model with an asymptotically linear test statistic, an analysis similar to~\Cref{fig:bound-comparison} may describe the performance along a one-dimensional line of parameter space corresponding to the direction of the test.

\section{CSE Application Workflow}
\label{sec:cse-application-workflow}

In this section, we give a high-level overview
of the workflow of applying CSE.

Before proceeding with CSE, we must begin with a bounded parameter region of interest $\Theta$.
In many cases, physical bounds on realistic outcomes and effect sizes will restrict $\Theta$. Alternatively, attention can be restricted to a bounded confidence set with the procedure of~\citet{berger1994p} (1994), at a minor cost to the net confidence level. Or, as a supplement to CSE analysis on a bounded region, loose upper bounds on the remainder of the space may be derivable with analytical techniques. 

Although the use of CSE is not limited 
to Type I Error,
for simplicity we restrict our discussion in this section to the case of
Type I Error and Family-Wise Error Rate (FWER) in the case of multiple testing. However, we note in passing that it is straightforward to extend these results 
to study FDR and the bias of bounded estimators.

A basic issue will arise as we move to multiple hypothesis testing: 
the FWER is not necessarily continuous 
across hypothesis boundaries. 
Fortunately, with $p$ hypotheses, 
the FWER function can be decomposed into $2^p$ smooth parts and thus divide-and-conquered.

We begin with null hypotheses $\mathcal{H}_j \subseteq \R^d$ for $j = 1,\ldots, p$.
Define $\Theta := \set{\theta : \theta \in \bigcup\limits_{j=1}^p \sH_j}$ to be the null
hypothesis space,
i.e. the space of parameters such that 
at least one of the null hypotheses is true.
The null hypotheses induce a natural partition of $\Theta$ into $2^p-1$ subsets with disjoint interiors where 
each subset completely resolves 
whether each null hypothesis is true.
Concretely, the partition is defined by $P(b)$
for some \emph{(null) configuration} $b \in \set{0,1}^{p} \setminus \set{\vec{0}}$ where
\begin{align}
    P(b)
    :=
    \bigcap\limits_{j : b_j = 0} \sH_j^c
    \cap
    \bigcap\limits_{j : b_j = 1} \sH_j
    \label{eq:partition:def}
\end{align}

Following this setup,
we define a few terminologies.

\begin{definition}[Tile and Platten]
\label{def:tile}
Suppose $\sH_j$ are null hypotheses for $j=1,\ldots, p$.
Let $\Theta$ be the null hypothesis space
induced by $\set{\sH_j}_{j=1}^p$.
A subset $H \subseteq \Theta$ is a
\emph{tile} if the interior of $H$ is a subset of $P(b)$
as in~\labelcref{eq:partition:def} for some configuration
$b \in \set{0,1}^{p} \setminus \set{\vec{0}}$.
Consider a collection of tiles $\set{H_i}_{i=1}^I$ 
with disjoint interiors that partition $\Theta$ and a collection of arbitrary
points $\set{\theta_i}_{i=1}^I \subseteq \Theta$ associated with each tile at the same index.
We say the pair $\set{H_i}_{i=1}^I, \set{\theta_i}_{i=1}^I$ 
is a \emph{platten}. 
\end{definition}

For every null configuration $b \in \set{0,1}^p \setminus \set{\vec{0}}$, 
the Type I Error to control FWER
at any $\theta \in P(b)$,
is given by $f_b(\theta) := \EEE_\theta\br{F_b(X)}$ where
$X \sim P_{\theta}$ and $F_b(x)$ is the indicator that a given (arbitrary) test with data $x$ rejects
for some $j$ where $b_j = 1$,
i.e. rejects at least one of the true null hypotheses 
in configuration $b$.

We give a sketch of the CSE application workflow.
Consider a platten $\set{H_i}_{i=1}^I$, $\set{\theta_i}_{i=1}^I$.
Let $b(i)$ be such that $H_i$ corresponds to 
the null configuration $b(i)$.
Suppose the user gathers some information
about the Type I Error at each $\theta_i$.
We apply~\Cref{thm:tilt-bound}
for each tile to get that
\begin{align*}
    \sup\limits_{v \in H_i-\theta_i}
    f_{b(i)}(\theta_i + v)
    &\leq
    \inf_{q\geq 1}
    \sup_{v \in H_i-\theta_i}
    U(\theta_i, v, q, f_{b(i)}(\theta_i))
\end{align*}
where $U$ is the Tilt-Bound~\labelcref{eq:tilt-bound:def}.
That is, we use the Tilewise Optimized Tilt-Bound to bound the 
worst-case Type I Error in each tile.
Combining the inference at each $\theta_i$
and the Tilewise Optimized Tilt-Bound,
we ``extend'' our inference to each tile.
Using a divide-and-conquer strategy to cover many tiles,
we again ``extend'' our inference to \emph{all of $\Theta$}.
In~\Cref{sec:validation,sec:calibration},
we give concrete methods utilizing this workflow.
Specifically,~\Cref{sec:validation}
constructs valid confidence intervals on
all of $\Theta$ by extending 
the confidence intervals at each $\theta_i$; 
\Cref{sec:calibration} constructs a critical
threshold for any arbitrary test that achieves
(average) level $\alpha$ on all of $\Theta$ by combining
calibrated thresholds gathered at each $\theta_i$.

\section{Validation Procedure}\label{sec:validation}

In this section, we describe our \emph{validation procedure}
that provides confidence bounds on the Type I Error
for any bounded null hypothesis space $\Theta$.
Concretely, given an arbitrary design $\sD$
and the family of distributions for the data
$\set{P_\theta : \theta \in \Theta}$,
we wish to construct random functions
$\pr{\hat{\ell}(\cdot), \hat{u}(\cdot)}$
such that for any $\delta > 0$,
\begin{align*}
    \forall \theta \in \Theta,\,
    \prob\pr{\hat{\ell}(\theta) \leq f(\theta)} \geq 1-\delta
    \text{ and } 
    \prob\pr{\hat{u}(\theta) \geq f(\theta)}
    \geq 1-\delta
\end{align*}
where $f(\theta)$ is the Type I Error of the design $\sD$ when data comes from $P_\theta$.
Note that we give a pointwise valid guarantee
for $\hat{u}$ rather than a uniform guarantee.
Under the frequentist framework, since 
we assume the existence of only one true parameter 
that generates the data,
it is sufficient to give a pointwise guarantee.
To simplify the discussion, we restrict 
our attention to the upper bound $\hat{u}(\cdot)$,
however, our methodology readily extends to 
the construction of $\hat{\ell}$ (see~\Cref{rem:extension-lower}). 

We now discuss the validation procedure in detail.
Following the workflow laid out in~\Cref{sec:cse-application-workflow},
we first discuss
in~\Cref{ssec:validation:point}
the simple method of constructing valid confidence bounds for a point-null. 
In~\Cref{ssec:validation:tile}, we extend this
guarantee at a point to a tile using CSE.
In~\Cref{ssec:validation:general},
we divide-and-conquer to construct our desired
upper bound function $\hat{u}(\cdot)$ to give guarantees
on all of $\Theta$.
To summarize the validation procedure,
we demonstrate it on the one-sided z-test
in~\Cref{ssec:validation:z-test}.


\subsection{Validation on a Single Point}
\label{ssec:validation:point}

Validation on a single point is the simplest case.
The task at hand is to construct an upper bound 
$\hat{u}$ given any $\delta > 0$
such that at a point $\theta_0$,
\begin{align*}
    \prob_{\theta_0}\pr{f(\theta_0) \leq \hat{u}}
    \geq
    1-\delta
\end{align*}
Although there are many choices for $\hat{u}$,
we use the Clopper-Pearson upper bound 
when $f(\theta)$ is the Type I Error function. (In the case of bounded estimates, this can be replaced with other inequalities such as Hoeffding's inequality \cite{hoeffding1994probability}; but we refrain from exploring this further). Given simulations $i=1,\ldots, N$,
let $B_i$ be the indicator that simulation $i$ (falsely) rejects.
Let $R = \sum\limits_{i=1}^N B_i$ be the number
of false rejections.
Then set the Clopper-Pearson upper bound
\begin{align}
    \hat{u}
    :=
    \Beta^{-1}(1-\delta, R + 1, N - R)
    \label{eq:clopper-pearson:def}
\end{align}

where $\Beta^{-1}(q, \alpha, \beta)$ is the
$q$th quantile of $\Beta(\alpha, \beta)$ distribution~\citep{clopper:1934}. An attractive feature of the Clopper-Pearson bound is that $\hat{u}$ is finite-sample valid.

For analysis of FDR or bias of a bounded estimator, the Clopper-pearson interval can be replaced with another finite-sample interval (such as Hoeffding confidence bounds.)

\subsection{Validation on a Tile}
\label{ssec:validation:tile}

We now discuss how to ``extend'' the confidence
bound from~\Cref{ssec:validation:point} to a tile.

Let $H$ be a tile (see~\Cref{def:tile}) with
an associated point $\theta_0$.
Suppose that $\hat{u}_0$ is the $(1-\delta)$
Clopper-Pearson upper bound~\labelcref{eq:clopper-pearson:def}.
Fix any displacement $v \in H - \theta_0$.
Using the fact that the Tilt-Bound
$a \mapsto U(\theta_0, v, q, a)$ is increasing,
which is immediate from the definition, we have that
\begin{align}
    \prob_{\theta_0}\pr{f(\theta_0+v) \leq U(\theta_0, v, q, \hat{u}_0)}
    &\geq
    \prob\pr{f(\theta_0) \leq \hat{u}_0}
    \geq
    1-\delta
    \label{eq:validation:tile:proof}
\end{align}
This shows that we may extend the upper bound function over H as
\begin{align}
    \hat{u}(\theta_0+v) 
    := 
    U(\theta_0, v, q, \hat{u}_0)
    \label{eq:validation:ub}
\end{align}
to achieve a pointwise valid confidence guarantee. 
As a conservative simplification we may consider the worst-case Tilt-Bound estimate, flattening the bound over $H$:
\begin{align*}
    \hat{u}(\cdot)
    :=
    \sup\limits_{v \in H-\theta_0}
    U(\theta_0, v, q, \hat{u}_0)
\end{align*}
Additionally, with the linearity condition of~\Cref{thm:tilt-bound-qcv},
we can compute this supremum easily for convex polytope
$H$ (see~\Cref{thm:tilt-bound-qcv}).
Finally, in either case, we may also 
minimize over $q \in [1,\infty]$ to get
a tighter bound for free
(see~\Cref{thm:tilt-bound-qcp} regarding computation).

In practice, we typically 
choose $\theta_0$ close to elements of $H$
(e.g. the center of $H$).
However, the theory does not enforce any
particular way of choosing $\theta_0$.
Also, as $H$ shrinks in size, 
we incur less extrapolation cost
from the Tilt-Bound, resulting in a tighter 
upper bound $\hat{u}$.

%
%
%
%
%
%
%
%

\subsection{Validation on a General Bounded Space}
\label{ssec:validation:general}

Given a general bounded null hypothesis space $\Theta$,
we begin with a platten $\set{H_i}_{i=1}^I$, $\set{\theta_i}_{i=1}^I$ (see~\Cref{def:tile}).
To construct an upper bound $\hat{u}(\cdot)$ for all of $\Theta$, we use the method described in~\Cref{ssec:validation:tile} to construct 
valid upper bounds $\hat{u}_i(\cdot)$ for each tile $H_i$
associated with $\theta_i$.
Then, we simply combine this collection into one function,
namely
\begin{align*}
    \hat{u}(\theta)
    :=
    \hat{u}_i(\theta)
    \quad
    \text{if $\theta \in H_i$}
\end{align*}
Note that by definition of platten, 
the interior of $H_i$ are disjoint so that
$\hat{u}(\theta)$ is well-defined whenever 
$\theta$ lies in the interior of some $H_i$.
For $\theta$ on the boundaries of possibly multiple $H_i$,
one may arbitrarily define $\hat{u}$ to take 
any one of the corresponding $\hat{u}_i$,
so long as the choice is not random.
Or, one could also define 
$\hat{u}(\theta) := \max\limits_{j : \theta \in H_j} \hat{u}_j(\theta)$, 
the maximum bound across all tiles containing $\theta$.
The validation procedure is summarized in~\Cref{alg:validation}.

\begin{algorithm}[t]
\caption{Validation Procedure}\label{alg:validation}
\begin{algorithmic}[1]
\State 
    Construct a platten $\set{H_i}_{i=1}^I$, $\set{\theta_i}_{i=1}^I$ for
    the (bounded) null hypothesis space 
    $\Theta$.
\State Construct a $(1-\delta)$ Clopper-Pearson 
    interval at $\theta_i$.
\State Use the Optimized Tilt-Bound to extend 
    the confidence bound at $\theta_i$ to all points in $H_i$.
\State Combine the tilewise confidence bounds into a single bound on $\Theta$.
\end{algorithmic}
\end{algorithm}

It is worth noting that $\hat{u}$ only depends on
simulations at finitely many points, namely $\set{\theta_i}_{i=1}^I$, so $\hat{u}$
is computable.
However, the implication is indeed that we have a pointwise
valid confidence bound on all of $\Theta$, not just
at the simulated points $\set{\theta_i}_{i=1}^I$.

In practice, the usefulness of $\hat{u}$ depends on the fineness of the tiles. 
A rougher partition will require less computation, but
will result in a more conservative $\hat{u}(\cdot)$. 
In any case, the confidence guarantee is valid in finite samples. 

\subsection{Validation for The Z-Test}%
\label{ssec:validation:z-test}

We now demonstrate the validation procedure 
for the one-sided z-test (see~\Cref{ssec:tilt-bound-normal}) for testing $H_0: \theta \leq 0$
at level $\alpha = 2.5\%$.
Traditionally, the null hypothesis space is $(-\infty, 0]$, however, to satisfy the boundedness
assumption, we limit our attention to $\Theta := [-1, 0]$.

\begin{figure}[t]
    \centering
    \begin{subfigure}[b]{0.49\textwidth}
    \centering
    \includegraphics[width=\textwidth]{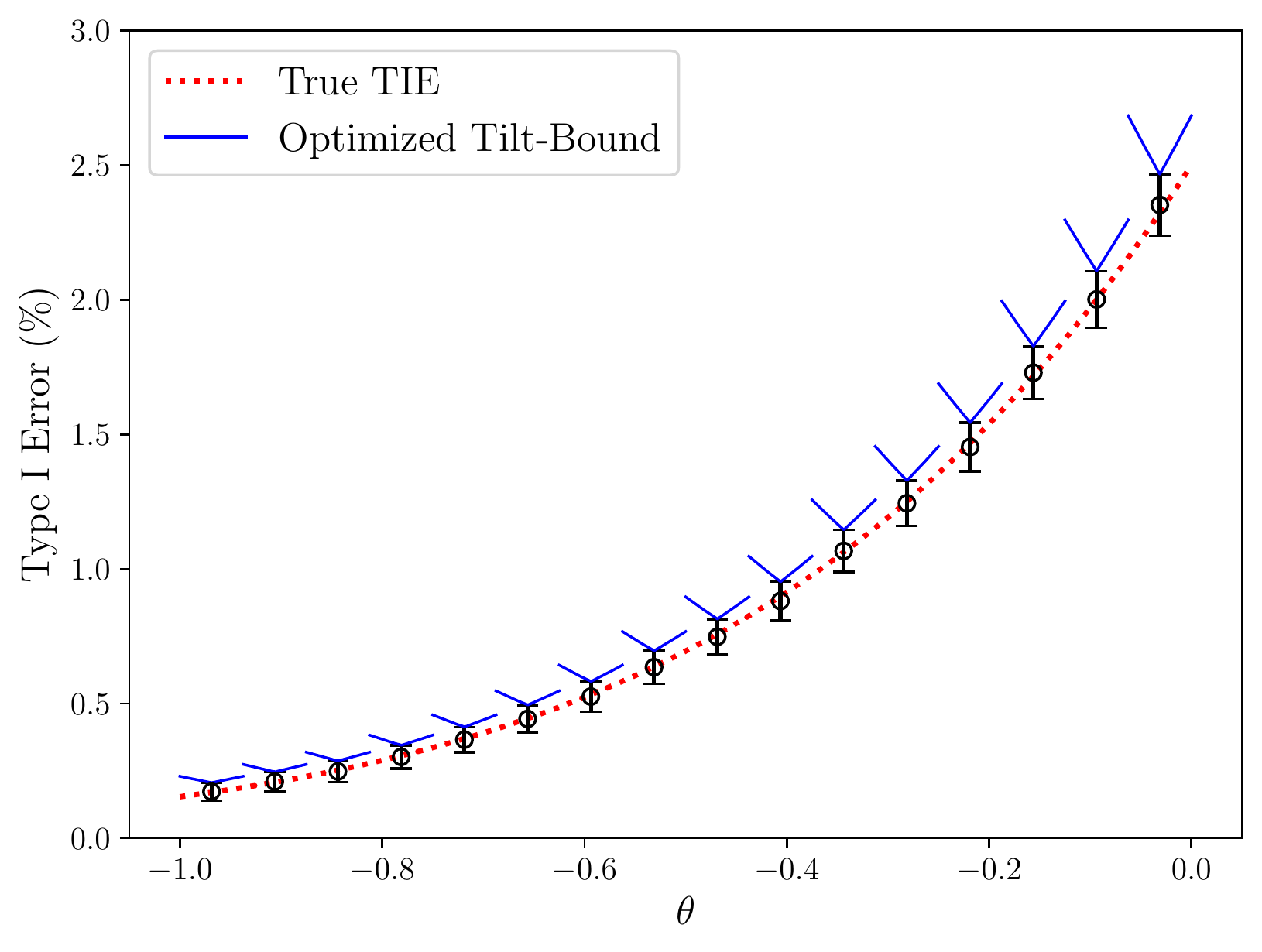}
    \caption{}
    \label{fig:z-test-tile-bound:16}
    \end{subfigure}
    \begin{subfigure}{0.49\textwidth}
    \centering
    \includegraphics[width=\textwidth]{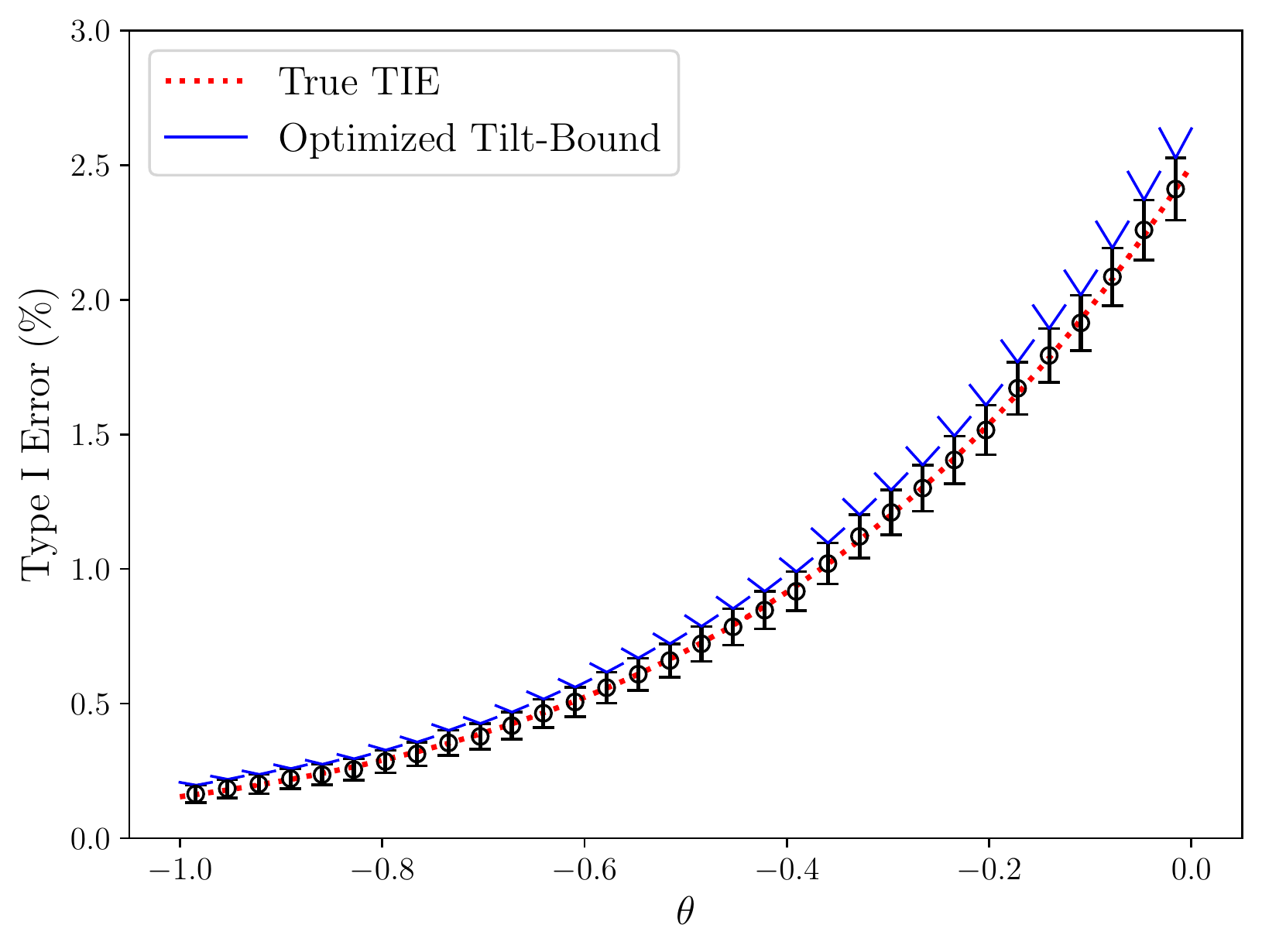}
    \caption{}
    \label{fig:z-test-tile-bound:32}
    \end{subfigure}
    \caption{Plot of the optimized Tilt-Bound 
    (\textbf{\textcolor{blue}{blue}} solid line)
    for the z-test
    Clopper-Pearson confidence intervals
    (\textbf{\textcolor{black}{black}} vertical bars)
    overlaid with the true Type I Error
    (\textbf{\textcolor{red}{red}} dotted line).
    \Cref{fig:z-test-tile-bound:16}
    uses 16 equally-spaced simulation points and 
    \Cref{fig:z-test-tile-bound:32} uses 32 points.
    This shows that with a finer gridding,
    the cost of the optimized Tilt-Bound quickly becomes negligible.
    }
    \label{fig:z-test-tile-bound}
\end{figure}

Following~\Cref{ssec:validation:general},
we choose equally-spaced points in the interval $(-1, 0)$, which naturally define
a platten for $\Theta$ where $\theta_i$ are taken
to be the centers of the tiles (intervals) $H_i$
with equal radius.
For each simulation point $\theta_i$,
we construct
$(1-\delta)$ Clopper-Pearson upper bounds, where $\delta \equiv 0.05$.
We then compute the upper bounds for each tile as in~\Cref{ssec:validation:tile}.

\Cref{fig:z-test-tile-bound} shows 
the validation results for the one-sided z-test
using 16 simulation points (\labelcref{fig:z-test-tile-bound:16}) and 32 simulation points (\labelcref{fig:z-test-tile-bound:32}).
Here, the Optimized Tilt-Bound refers to
the upper bound $\hat{u}$ as in~\labelcref{eq:validation:ub} optimized over $q\in [1,\infty]$.
We overlay the true Type I Error for comparison
and include the Clopper-Pearson intervals
at each simulation point.
Notice that the bounds are all above the true
Type I Error curve, lying close to the curve. 
Depending on the tile, the flatness of the bound
adapts as well: the smaller the $\theta$,
the flatter the bound.
This demonstrates that the Tilt-Bound
is using the ``curvature information'' of the 
Normal location family quite well
and accurately depicts the area of difficult 
estimation, namely the area close to the null hypothesis boundary.
By comparing~\Cref{fig:z-test-tile-bound:16,fig:z-test-tile-bound:32}, we see that the cost of 
the Tilt-Bound becomes neglible as 
the platten becomes finer, to the point where
the cost of Clopper-Pearson bounds dominate.
Note, however, that the Clopper-Pearson bounds 
can be tightened by increasing the number of simulations.

\section{Calibration Procedure}\label{sec:calibration}

The \emph{calibration procedure} 
provides a level-$\alpha$ expected Type I Error control guarantee at every point of the bounded null hypothesis region $\Theta$. 
In the sense that a confidence interval's guarantee is valid 
averaged over possible observations of the data,
the calibration guarantee is valid averaged
over the possible threshold choices that result from the process. 
In general, the actual expected Type I Error is smaller than the nominal guarantee, and can be much smaller if the simulation scale is not sufficiently large. But the guarantee is still valid in this case, and the sponsor is incentivized to provide more comprehensive simulations to increase power. The fail-safe properties of calibration may be highly desirable for regulatory applications. 
These properties should be compared against the \textit{validation} procedure described earlier in~\Cref{sec:validation}
in which the design's critical threshold is fixed, but randomness is present in the guarantee level so that a target could be randomly overshot. Moreover, the validation confidence bound only 
holds with probability $1-\delta$. 

Given an arbitrary design $\sD$,
the family of distribution for the data
$\set{P_\theta : \theta \in \Theta}$,
and the level $\alpha$,
we wish to construct a (random) threshold $\hat{\lambda}$
such that
\begin{align*}
    \forall \theta \in \Theta,\,
    \EEE\br{f_{\hat{\lambda}}(\theta)} \leq \alpha
\end{align*}
where $f_\lambda(\theta)$ is the Type I Error 
of the design $\sD$ using the critical threshold $\lambda$, when the data comes from $P_\theta$.
To be more precise, we assume $f_\lambda(\theta)$ takes on the form
\begin{align*}
    f_{\lambda}(\theta) := \prob_\theta\pr{S(X) < \lambda}
\end{align*}
where $S(X)$ is the test statistic of $\sD$
and $X \sim P_\theta$.
As an example, for the one-sided z-test, 
it is natural to consider $S(X) := -X$
so that $S(X) < \lambda$ corresponds 
to an upper-tail rejection.
Equivalently, we could use the p-value
$S(X) := 1 - \Phi(X)$
where $\Phi$ is the standard normal CDF.

As in~\Cref{sec:validation},
we first discuss how to calibrate for a single point in~\Cref{ssec:calibration:point}.
\Cref{ssec:calibration:tile} extends
this guarantee at a point to a tile using CSE.
In~\Cref{ssec:calibration:general},
we divide-and-conquer to construct a single
critical threshold that achieves level $\alpha$
on all of $\Theta$.
Finally, we apply the calibration procedure
on the one-sided z-test in~\Cref{ssec:calibration:z-test}.

\subsection{Calibration on a Single Point}
\label{ssec:calibration:point}

We begin with the calibration procedure
on a single null point $\theta_0$.

Suppose $N$ simulations are performed at $\theta_0$ 
so that we collect $S_1,\ldots, S_N$ i.i.d.
samples of the test statistic under $P_{\theta_0}$.
Let 
\begin{align}
    \hat{\lambda}^\star := S_{(\floor{(N+1)\alpha})}
    \label{eq:calibrated-threshold}
\end{align}
where $S_{(i)}$ is the $i$th order statistic among
$S_1,\ldots, S_N$.
Then, by an exchangeability argument, 
\Cref{thm:calibration} shows that
\begin{align*}
    \EEE_{\theta_0}\br{f_{\hat{\lambda}^\star}(\theta_0)} 
    \leq 
    \frac{\floor{(N+1)\alpha}}{N+1} 
    \leq 
    \alpha
\end{align*}
Hence, this result allows us 
to \emph{target} a Type I Error upper bound 
at a given point 
for any level $\alpha$ by choosing 
the appropriate quantile of the test statistics.

\Cref{thm:calibration} tells us more.
First, it lower bounds 
the average Type I Error in~\labelcref{eq:calibration:tight-mean}.
Moreover,~\labelcref{eq:calibration:tight-var}
shows that the variance of the procedure consists 
of a component that decays at rate $O\pr{\frac{1}{N}}$
and a second component due to jumps in the distribution. 
This second component vanishes in the case of a continuous test statistic $S(X)$ and can thus be removed by randomizing the test - in general, this may be achieved by adjoining an independent uniform random variable to the sample space, and re-defining the test statistic as $(S,u)$ and the threshold as $(\lambda,c)$ to be compared in the lexicographic order. 
As a result, we also have concentration
of $f(\hat{\lambda}^\star)$. 
Thus, power is well-maintained for sensible tests without heavy discretization.

\begin{theorem}[Pointwise calibration]\label{thm:calibration}
Let $S_1,\ldots, S_N$ be any i.i.d. random variables.
Fix any $\alpha \in \br{\frac{1}{N+1}, \frac{N}{N+1}}$.
Define the following functions:
\begin{align*}
    f_+(\lambda) &:= \prob\pr{S \leq \lambda} \\
    f(\lambda) &:= \prob\pr{S < \lambda} \\
    \Delta f(\lambda) &:= f_+(\lambda) - f(\lambda)
\end{align*}
Finally, let
$\hat{\lambda}^\star$ as in~\labelcref{eq:calibrated-threshold} and
$\delta_{N, \alpha} := \EEE\br{\Delta f(\hat{\lambda}^\star)}$.
Then,
\begin{align}
    \frac{\floor{(N+1)\alpha}}{N+1} 
    - \delta_{N, \alpha}
    \leq
    \EEE\br{f(\hat{\lambda}^\star)} 
    \leq 
    \frac{\floor{(N+1)\alpha}}{N+1} 
    \label{eq:calibration:tight-mean}
\end{align}
Moreover,
\begin{align}
    \var{f(\hat{\lambda}^\star)}
    &\leq
    O\pr{\frac{1}{N}}
    +
    \delta_{N,\alpha}
    \pr{\frac{2\floor{(N+1)\alpha}}{N+1} - \delta_{N,\alpha}}
    \label{eq:calibration:tight-var}
\end{align}
\end{theorem}

\subsection{Calibration on a Tile}
\label{ssec:calibration:tile}

We now discuss calibration on a tile.

Let $H$ be a tile (see~\Cref{def:tile}) with an associated point $\theta_0$.
We wish to construct $\hat{\lambda}^\star$ such that
\begin{align*}
    \sup\limits_{v \in H-\theta_0} 
    \EEE\br{f_{\hat{\lambda}^\star}(\theta_0+v)} 
    \leq
    \alpha
\end{align*}
Note that for any random rejection rule $\hat{\lambda}$, 
the Tilt-Bound guarantee in~\Cref{thm:tilt-bound} holds so that
\begin{align}
    \EEE\br{f_{\hat{\lambda}}(\theta_0+v)}
    &\leq
    U(\theta_0, v, q, \EEE\br{f_{\hat{\lambda}}(\theta_0)})
    \label{eq:tilt-bound-bwd}
\end{align}
for any $q \geq 1$.
This is because by Fubini's Theorem, we may view
\begin{align*}
    \EEE\br{f_{\hat{\lambda}}(\theta)}
    &=
    \EEE_{X \sim P_\theta} \EEE_{\hat{\lambda}}\br{\indic_{S(X) < \hat{\lambda}}}
    =
    \EEE_{X \sim P_\theta} \br{G(X)}
\end{align*}
where $G(x) := \prob\pr{S(x) < \hat{\lambda}} \in [0,1]$.
Hence, we may apply~\Cref{thm:tilt-bound} 
with $G$ as the measurable function to integrate.

By back-solving~\labelcref{eq:tilt-bound-bwd}
for $\EEE\br{f_{\hat{\lambda}}(\theta_0)}$, we get that
\begin{align}
    \sup\limits_{v \in H-\theta_0}
    U\pr{\theta_0, v, q, \EEE\br{f_{\hat{\lambda}}(\theta_0)}}
    \leq \alpha
    \iff
    \EEE\br{f_{\hat{\lambda}}(\theta_0)}
    &\leq
    \inf\limits_{v \in H}
    U^{-1}(\theta_0, v, q, \alpha)
    \label{eq:tile-calibration}
\end{align}
where $U^{-1}$ is the \emph{Inverted Tilt-Bound} defined by
\begin{align}
    U^{-1}(\theta_0, v, q, \alpha) 
    &= \br{
    \alpha \exp\pr{
    -\frac{\psi(\theta_0, v, q)}{q} + \psi(\theta_0, v, 1)
    }}^{\frac{q}{q-1}}
    \label{eq:tilt-bound-inv}
\end{align}
Hence, the bound in~\labelcref{eq:tile-calibration}
can be thought of as a target level at $\theta_0$ to ensure the test 
will be level $\alpha$ on all of $H$.
Note that the bound in~\labelcref{eq:tile-calibration}
can be maximized over $q \in [1,\infty]$ to get the \emph{least}
conservative target at $\theta_0$.
So, we have reduced the problem to constructing $\hat{\lambda}^\star$ such that
\begin{align}
    \EEE\br{f_{\hat{\lambda}^\star}(\theta_0)}
    &\leq
    \sup\limits_{q\in [1,\infty]}
    \inf\limits_{v \in H}
    U^{-1}(\theta_0, v, q, \alpha)
    =: \alpha'
    \label{eq:tile-calibration-final}
\end{align}
By a similar argument as in~\Cref{thm:tilt-bound-qcv}, 
as soon as $H$ is a convex polytope and the linearity 
condition is satisfied, the infimum reduces to a finite minimum.
And similar to~\Cref{thm:tilt-bound-qcp},
the supremum is a one-dimensional quasi-concave maximization problem.
Together, the bound $\alpha'$ in~\labelcref{eq:tile-calibration-final} 
can be easily computed.
To construct $\hat{\lambda}^\star$, we simply use~\Cref{ssec:calibration:point}
with ``$\alpha$'' as $\alpha'$.
Hence, by relying on the Tilt-Bound to find a target level for $\theta_0$ and
simulating only at $\theta_0$,
we have successfully extended the Type I Error control to all of $H$, that is,
we have constructed a computable $\hat{\lambda}^\star$ such that
\begin{align*}
    \sup\limits_{v \in H-\theta_0} 
    \EEE\br{f_{\hat{\lambda}^\star}(\theta_0+v)} 
    \leq
    \alpha
\end{align*}

\subsection{Calibration on a General Bounded Space}
\label{ssec:calibration:general}

Given a general bounded null hypothesis space $\Theta$,
we begin with a platten $\set{H_i}_{i=1}^I$, $\set{\theta_i}_{i=1}^I$
(see~\Cref{def:tile}).
For each tile $H_i$, 
we first perform tilewise calibration as described in~\Cref{ssec:calibration:tile} 
to obtain calibrated thresholds $\hat{\lambda}_i^\star$ 
that achieves level $\alpha$ within each $H_i$. It is noteworthy that these guarantees permit correlating samples or re-using the simulation RNG across different values of $\theta_i$ so long as the simulations corresponding to each individual tile $H_i$ remain i.i.d.. Correlating the simulations offers not only computational savings but also smooths out the estimated Type I Error surface, which reduces the procedure's conservatism. After performing these calibrations, we may select the most conservative threshold over of the tiles,
\begin{align}
\hat{\lambda}^{\star} := \min\limits_{i=1,\ldots,I} \hat{\lambda}_i^\star
\label{eq:calibrated-threshold-final}
\end{align}
Then, since the rejection sets $\set{S(X) < \lambda}$ is monotonic in $\lambda$,
\begin{align*}
    \sup\limits_{\theta \in \Theta}
    \EEE\br{f_{\hat{\lambda}^{\star}}(\theta)}
    &=
    \max\limits_{i=1,\ldots, I}
    \sup\limits_{\theta \in H_i}
    \EEE\br{f_{\hat{\lambda}^{\star}}(\theta)}
    \leq
    \max\limits_{i=1,\ldots, I}
    \sup\limits_{\theta \in H_i}
    \EEE\br{f_{\hat{\lambda}^\star_i}(\theta)}
    \leq
    \alpha
\end{align*}
To summarize,~\Cref{alg:calibration} describes the full calibration procedure.
\begin{algorithm}[t]
\caption{Calibration Procedure}\label{alg:calibration}
\begin{algorithmic}[1]
\State Construct a platten $\set{H_i}_{i=1}^I$, $\set{\theta_i}_{i=1}^I$ for the (bounded) null hypothesis space $\Theta$.
\State For each tile $H_i$ and simulation point $\theta_i$, 
construct $\hat{\lambda}^\star_i$~\labelcref{eq:calibrated-threshold}
that satisfies~\labelcref{eq:tile-calibration-final}.
\State Select the most conservative threshold $\hat{\lambda}^{\star} := \min\limits_{i=1,\ldots,I} \hat{\lambda}^\star_i$.
\end{algorithmic}
\end{algorithm}

\subsection{Calibration on The Z-Test}
\label{ssec:calibration:z-test}

In this section, we apply the calibration procedure on the one-sided z-test.

\begin{figure}[t]
    \centering
    \includegraphics[width=0.8\textwidth]{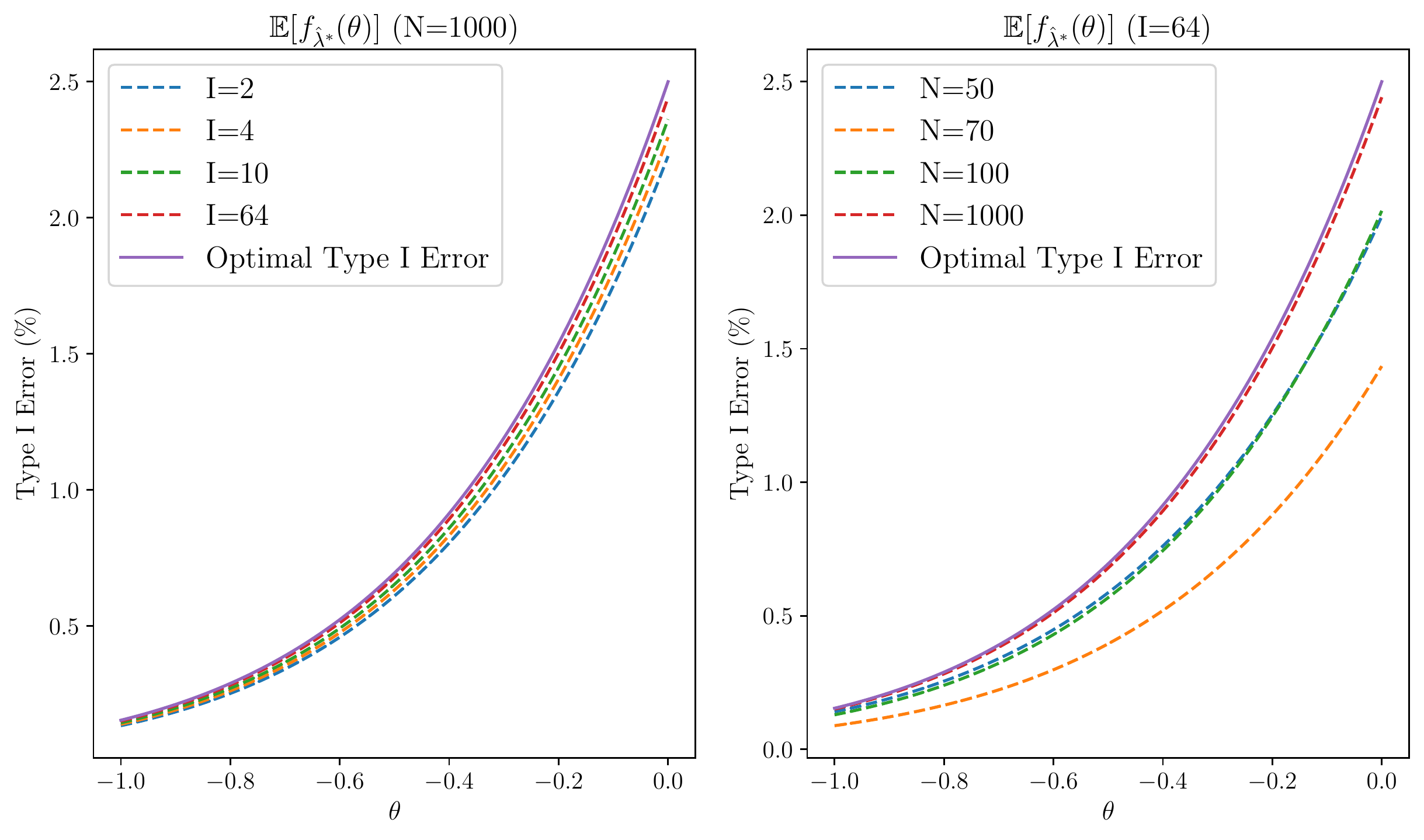}
    \caption{%
    We consider the one-sided z-test on $\Theta \equiv [-1,0]$.
    Both plots show a comparison of the average Type I Error  
    $\EEE\br{f_{\hat{\lambda}^\star}(\theta)}$ 
    using the calibrated threshold $\hat{\lambda}^\star$ as in~\labelcref{eq:calibrated-threshold-final}. It is assumed that correlated simulations will be used generate the z-score of dataset $i,k$ by translating a shared error $Z_k \sim N(0,1)$ by $\theta_i$. The curves shown for $\EEE\br{f_{\hat{\lambda}^\star}(\theta)}$ are derived analytically.
    The left fixes the calibration simulation size $N = 1000$
    while varying the number of tiles.
    Conversely, the right fixes the number of tiles $I = 64$
    while varying the number of simulations per-tile. Together they show that with greater computational power, nearly-optimal Type I Error can be achieved.
    }
    \label{fig:calibration-z-test}
\end{figure}

~\Cref{fig:calibration-z-test}, shows the Type I Error performance of the calibration procedure(see~\Cref{ssec:tilt-bound-normal}).
We plot the average Type I Error 
$\theta \mapsto \EEE\br{f_{\hat{\lambda}^\star}(\theta)}$
where $\hat{\lambda}^\star$ is constructed as in~\labelcref{eq:calibrated-threshold-final}.
We also show the optimal Type I Error, which is the usual curve
$\theta \mapsto 1 - \Phi(z_{1-\alpha} - \theta)$.
In the left plot, the calibration simulation size is fixed to $N = 1000$ and the number of tiles $I$ is varied.
Conversely, the right plot fixes the number of tiles $I = 64$
and varies the calibration simulation size. Both curves show close-to-optimal Type I Error performance with large $I$ and $N$. In the right plot the performance is not strictly increasing in $N$; this is simply due to discretization error in equation ~\labelcref{eq:calibrated-threshold} which occurs when $\alpha$ is not a multiple of $\frac{1}{N+1}$. For $\alpha = 2.5\%$, choosing N divisible by 200 is convenient to keep the discretization effect small.

\section{From Calibration to Confidence Sets and Intervals}\label{sec:confidence-set}

The calibrations discussed earlier in~\Cref{sec:calibration} 
can be used to form confidence regions and confidence intervals for parameters and estimands of interest, such as a treatment effect $p_1(\theta) - p_0(\theta)$. Below are the steps to construct a $(1-\alpha)$ confidence upper-bound on an inference target $e(\theta)$. To form a two-sided interval, one may simply intersect an upper and a lower interval, each at level $1-\alpha/2$.

We describe the algorithm for constructing an upper confidence interval:
\begin{enumerate}[label=\textbf{\arabic*}:]
    \item Select test statistic $s_i$ for the inference target $e(\theta)$. If desired, one may choose a different test statistic for each tile $H_i$. 
    \item Compute the tile-specific calibrations of~\Cref{ssec:calibration:tile}, resulting in critical values $\hat{\lambda}_i(\alpha)$.
    \item Form a multi-dimensional confidence region $C_{1-\alpha}$ within the domain $\Theta$, as a union of the non-rejected tiles:
    $$C_{1-\alpha} := \bigcup \limits_{i: s_i \geq \hat{\lambda}_i(\alpha)} H_i$$
    \item Form the confidence set $C^e_{1-\alpha}$ from the image of $C_{1-\alpha}$ under the function $e$, defined as 
    \begin{align*}
        C^e_{1-\alpha} := \set{e(\theta) : \theta \in C_{1-\alpha}}
    \end{align*}
\end{enumerate}

This construction trivially attains coverage of at least $1-\alpha$. 
The resulting confidence sets are conservative in general, but if an asymptotically linear test statistic is used on an asymptotically Gaussian model, the cost may be small. 
The construction could perhaps be improved by developing higher-order test corrections to cause $C_{1-\alpha}$ to appear like a level set of $e(\theta)$ (thus reducing loss), or through the use of an additional re-calibration. We leave these topics to future work.


\section{Handling CSE with Adaptive Procedures} \label{sec:cse-for-adaptive-design}

In this section we discuss how our theory developed for exponential families applies to a wide class of designs and outcome mechanisms. In particular, we discuss

\begin{enumerate}[label=(\arabic*)]
    \item \label{itm:adc} \textbf{Adaptive Data Collection}: a design where the number of data points to be collected follows a pre-specified plan.
    \item \label{itm:ac} \textbf{Administrative Censoring}: the right-censoring of survival data that occurs at time of analysis or study conclusion.
    \item \label{itm:lv} \textbf{Latent Variables}: outcome data with multiple states such as an HMM, or multi-step randomness such as a random frailty.
\end{enumerate}


To apply CSE to problems with adaptive sampling or stopping elements, we take the approach of embedding the adaptive model within a larger i.i.d. model. 
That is, in the case of an adaptive design with filtration 
$\mathcal{F}_t$ with maximum time (or sample size) $T$, we may trivially embed $\mathcal{F}_t$ in a finer $\sigma$-algebra $\mathcal{F}$ with i.i.d. sampling structure and possibly any other independent information.
Then, the simple structure of $\sF$ 
allows application of the Tilt-Bound to $\sF$-measurable tests. 
Because $\sF$ is finer, 
our theory immediately goes through 
for any tests which are $\sF_t$-adapted. 

For example, consider analyzing a one-arm trial with exponential outcomes with statistical parameter $\lambda$, with possible adaptive stopping and administrative censoring at time $t$. 
The filtration is $\mathcal{F}_t = \sigma(X_i \mathbbm{1}_{\{X_i \leq s\}}, \, 1 \leq i \leq n, \, s \leq t)$ where $X_i \sim \Exp(\lambda)$. 
We may embed this model in the larger $\mathcal{F} = \sigma(X_i, \, 1 \leq i \leq n)$, and perform simulations of i.i.d. exponentials. 
We must, of course, take care that the simulation's sequential rejection functions are correctly implemented and do not use any information past time $t$. The resulting Tilt-Bound on the Type I Error or other metric over the $\sigma$-algebra $\sF$ thus directly implies a bound under $\mathcal{F}_\tau$
for any stopping time $\tau$.

Next, consider the case of an adaptive sampling of binomial arms, such as in an analysis of the multi-arm bandit problem with $K$ arms, each with $\Bern(p_k)$ outcomes. 
In this case, the maximum possible number of samples that could be taken from each arm is $N$. 
We may embed this model in a $\sigma$-algebra $\sF$ that contains a total of $NK$ independent data points with $X_{nk} \sim \Bern(p_k)$. 
In this case, the Tilt-Bound can be applied to a matrix of $N$ samples from each of $K$ independent Bernoullis.

Note that in~\labelcref{itm:adc}, 
$F(X)$ must accord with the decision rule and prospective sampling 
decisions up to time $\tau$, 
and not depend on data beyond time $\tau$. 
In~\labelcref{itm:ac}, $F(X)$ must further not depend on the specific 
data values that are larger than the censoring times $\gamma_{i}$, 
beyond knowledge that $\indic\pr{s_i > \gamma_i}$ 
where $s_i$ is the $i$'th survival time. 
For~\labelcref{itm:lv}, $F(X)$ must not depend on the value of the 
latent variable (whether the unknown Hidden Markov Model (HMM) 
state or unknown individual patient hazard) 
beyond what can be determined by the ``visible" outcome data. 




\subsection{Calibration of Adaptive Designs with Interim Stopping for Efficacy}
\label{appendix:adaptive-calibration}

For calibration of adaptive designs with multiple interim analyses, defining the monotone family of designs indexed by $\lambda$ may require some care. It is tempting to define $\lambda$ as a global rejection-threshold across interim analyses, such as a critical threshold for the Bayesian posterior probability of efficacy for each treatment arm. 
This approach works for single-hypothesis studies, but in adaptive multi-arm designs it sometimes breaks the monotonicity condition required for calibration, because failing to reject one arm might cause a different arm to be rejected in the future. To avoid this issue, the design in ~\Cref{ssec:lewis} tunes the Bayesian posterior threshold only at the final analysis, rather than at all analyses simultaneously.


In practice, it may be sufficient to index $\lambda$ as the critical rejection threshold of the final analysis,
but in interest of absolute rigor we remark that it is possible for this calibration to fail if the simulations at some $\theta_i$ reject with probability greater than 2.5\% before the final analysis. 
This issue does technically have a
general, robust solution: the indexed family of designs can be expanded by removing the mass of the rejection set in order, from the largest time to the smallest time. To achieve this formally, one would define $\lambda$ according to Siegmund's ordering~\citep{lai2006confidence}, or equivalently, the lexicographic order on $(\tau, S_\tau)$, where $\tau$ is the stopping time and $S_\tau$ is a rejection statistic, such that a smaller value of $S_t$ should favor rejection. This extended definition of $\lambda$ embeds the calibration of the final analysis in a rigorous way and ensures the viability of the overall Type I Error proof.

However in practice, if a calibration eliminates an intended final analysis, this occurrence may indicate poor setup of the design or insufficient simulations. In this case, the practitioner should be empowered to reconsider their design and simulation setup. Loss of formal Type I Error control is preferable to using a poor or poorly-calibrated design.

\section{Application of CSE} \label{sec:application-examples}

We discuss two main applications of CSE.
\Cref{ssec:berry} studies a Bayesian basket trial from~\citet{berry2013bayesian}
where we apply the validation procedure to achieve a
tight upper bound estimate of the Type I Error surface.
\Cref{ssec:lewis} studies a complex Phase II/III selection design
where we apply the calibration procedure to search for
the critical threshold of the test that achieves level $\alpha$.

\subsection{Validation on a Bayesian Basket Trial}%
\label{ssec:berry}

\begin{figure}[t]
    \centering
    \includegraphics[width=0.9\textwidth]{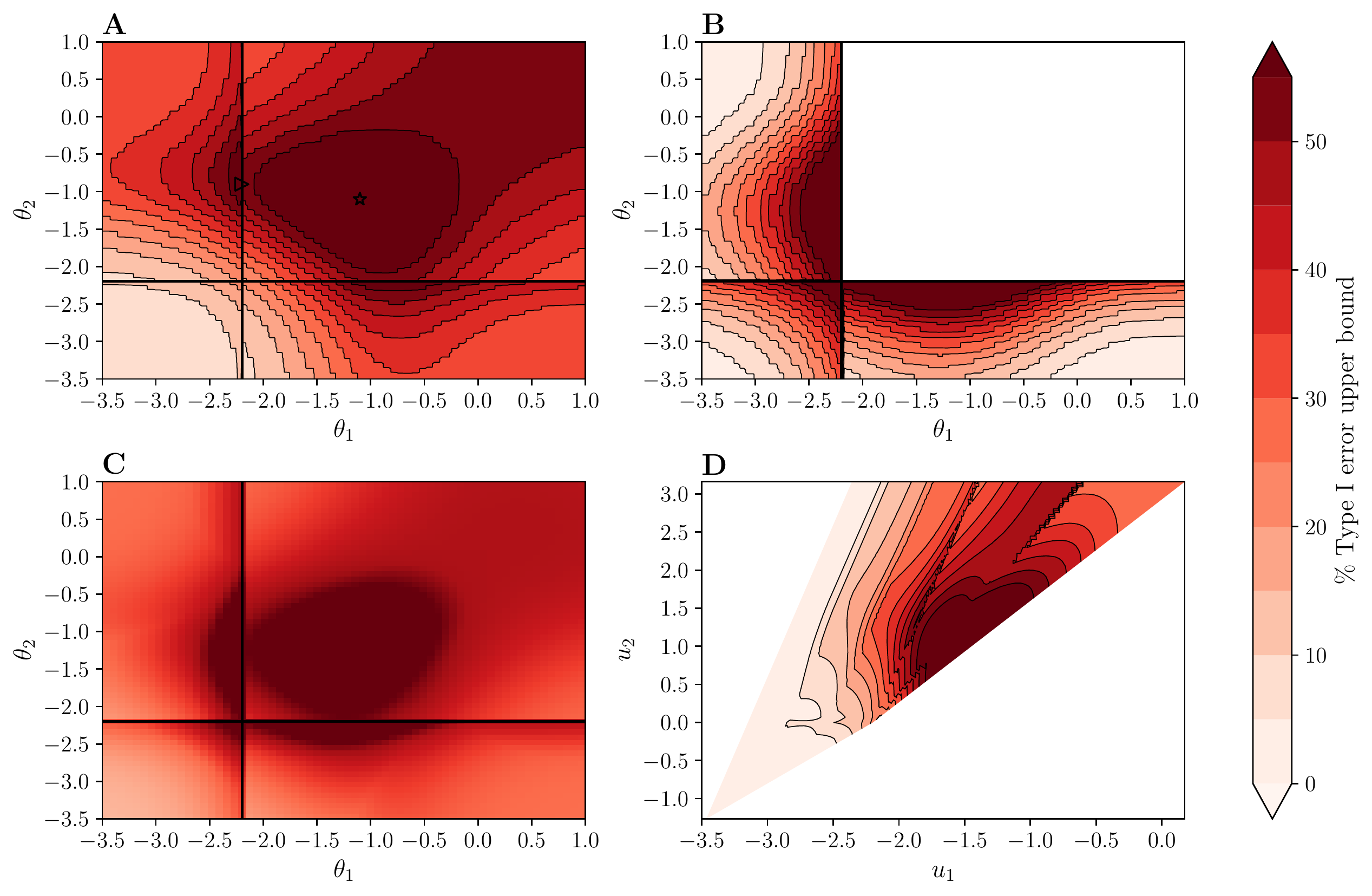}
    \caption{In plots \textbf{A}, \textbf{B} and \textbf{C}, the dark black lines indicate the boundary between the null space and the alternative space for each plotted parameter. \textbf{A)} The optimal Tilt-Bound as a function of $\theta_1$ and $\theta_2$ when $\theta_3= \theta_4= \theta_{critical}$. The open triangle and open star indicate the two peaks in Type I Error discussed in the text. 
    \textbf{B)} The optimal Tilt-Bound as function of $\theta_1$ and $\theta_2$ with $\theta_3$ and $\theta_4$ fixed at 
    the global maximum. 
    Note that the white space in the figure indicates the parameter sets would fall in the alternative space for all arms, and therefore, the Tilt-Bound is not meaningful. 
    \textbf{C)} For each value of $(\theta_1,\theta_2)$, the error surface shown is the worst-case Type I Error over $(\theta_3, \theta_4)$. \textbf{D)} For each $u_1=(\theta_1+ \theta_2+ \theta_3+ \theta_4)/4$ and $u_2=\max \limits_i (\theta_i-\theta_{0i})$, the error surface shown is the worst-case Type I Error over the remaining 2 dimensions}
    \label{fig:berry-output}
\end{figure}

In this section,
we examine the Type I Error (Family-Wise Error Rate)
of a Bayesian basket trial, 
modeled after the design of~\citet{berry2013bayesian}. 
Using the validation procedure described in~\Cref{sec:validation},
we obtain a tight upper bound estimate of the 
Type I Error using the Tilt-Bound.
The resulting landscape in~\Cref{fig:berry-output} 
is complex and non-monotonic. A total of 7.34 trillion simulations 
of the trial were performed, taking 4 hours on 
a single Nvidia V100 GPU.

The model we consider examines the effectiveness of a single treatment applied to different groups of patients distinguished by a biomarker. The model has 4 arms indexed $i = 1,\ldots,4$ and each patient has a binary outcome indicating treatment success or failure with probability $p_i$. 
Every arm has $n_i = 35$ patients. 
We model the outcome for each arm as:
\begin{equation}
    y_i \sim \Binom(n_i,p_i)
\end{equation}
The hierarchical model allows for data-derived borrowing between the arms and is described in the log-odds space as:
\begin{align*}
\theta_i \lvert \mu, \sigma^2 &\sim \Normal(\mu, \sigma^2) \\
\mu &\sim \Normal(-1.34, 100) \\ 
\sigma^2 &\sim \Gamma^{-1}(0.0005, 0.000005) 
\end{align*}
where $\theta_i$ determines $p_i$ according to:
\begin{equation}
\theta_i = \theta_{0i} + \logit(p_i)
\end{equation}
where $\logit(p) = \log(p / (1-p))$ and
$\theta_{0i}$ is a pre-determined constant
indicating the effectiveness of the standard of care for arm $i$. 
Our use of the inverse-Gamma prior is for historical reasons, 
and may no longer be recommended (see~\citep{cunanan2019variance}
for suggestions of alternative priors). 
This hierarchical model borrows between the arms with greater posterior certainty if the four groups perform very similarly. 
Rejection for arm $i$ is made according 
to the posterior probability using the following rule:
\begin{equation}
    \prob\pr{p_i > p_{0i} \vert \mathbf{y}} > 85\%
\end{equation}
where $\mathbf{y} := (y_1, y_2, y_3, y_4)$
is the vector of the binomial responses $y_i$.

We consider $\theta_i \in [-3.5, 1.0]$ for every $i$
and show four different meaningful portions of 
the resulting Tilt-Bound surface in~\Cref{fig:berry-output}.
The Tilt-Bound is as high as 60\% for some parameter combinations,
which is not surprising given that the design was not optimized 
to control the Type I Error. 
Nevertheless, the Tilt-Bound is well-controlled at 
the global null point where $\theta_i = \theta_{oi}$ for all arms. This is consistent with the results of~\citet{berry2013bayesian}, which calibrated these Bayesian analysis settings to ensure a low Type I Error at the global null point for a slightly more complicated version of this design.

The Tilt-Bound shows significant multi-modality with high error both near the null hypothesis boundary and at another peak further into the alternative space of $\theta_1$ and $\theta_2$. This multi-modality is due solely to the sharing effects (as the design is not adaptive). The open triangle and open star on~\Cref{fig:berry-output}A indicate the peaks in the Tilt-Bound and correspond, respectively, to the ``One Nugget" and ``2 Null, 2 Alternative" simulations from~\citet{berry2013bayesian}.
The first peak (indicated by an open triangle in~\Cref{fig:berry-output}A) occurs when the three variables, $\theta_1$, $\theta_3$ and $\theta_4$, are at their respective null hypothesis boundaries and the final variable, $\theta_2$ is deep into its alternative space. 
The sharing effect from the alternative-space variable \emph{pulls} estimates of the three null-space variables across the rejection threshold, resulting in high Type I Error.
Similarly, the second peak (indicated by an open star in~\Cref{fig:berry-output}A) occurs when the un-plotted variables, $\theta_3$ and $\theta_4$, lie on their respective null hypothesis boundaries and the hierarchical model pulls their estimates towards the plotted alternative-space variables $\theta_1$ and $\theta_2$. 

\subsection{Calibration on a Phase II/III Selection Design}\label{ssec:lewis}

In this section, we calibrate a Phase II/III Bayesian selection design which was suggested as a case study by an FDA official in personal communication.
We wish to perform the calibration procedure 
as in~\Cref{sec:calibration}
to obtain a level $\alpha$ test. Apart from calibration, the design is not otherwise optimized for performance. For general discussion on performing calibration of designs with adaptive stopping, see~\Cref{appendix:adaptive-calibration}. 

This Phase II/III selection design has two stages: the first stage has 3 treatment arms and 1 control arm, and it may select a treatment for continuation against control in the second stage. Outcomes are assumed $\Bern(p_i)$ for $i=0,\ldots, 3$, 
with $i = 0$ representing the control arm, and trial decisions are informed by the Bayesian hierarchical model described in~\Cref{ssec:berry} using data from all arms. At each of 3 interim analyses in the first stage, decision options include whether to stop for futility, to drop one or more poor-performing treatments, or to select a treatment for acceleration to the second stage. The second stage has one interim and one final analysis. The total number of patients across all arms and stages is at most 800 with at most 350 in any single arm. A full description is given in~\Cref{sec:appendix-d}.

We now restrict our attention to a bounded region of parameter space: the 4-dimensional cube $\Theta = [-1,1]^4$,
where $\theta \in \Theta$ assigns
$\logit(p_i)$ for each arm $i$ (so that $p_i$ ranges approx. [27\%, 73\%]). Each treatment arm for $i$ in $1, 2, 3$ has a null hypothesis denoted $H_i$, with null region given by
\[
\Theta_i = \set{\theta \in [-1,1]^4 : \theta_i \leq \theta_0}
\]

\begin{figure}
    \centering
    \includegraphics[width=12cm]{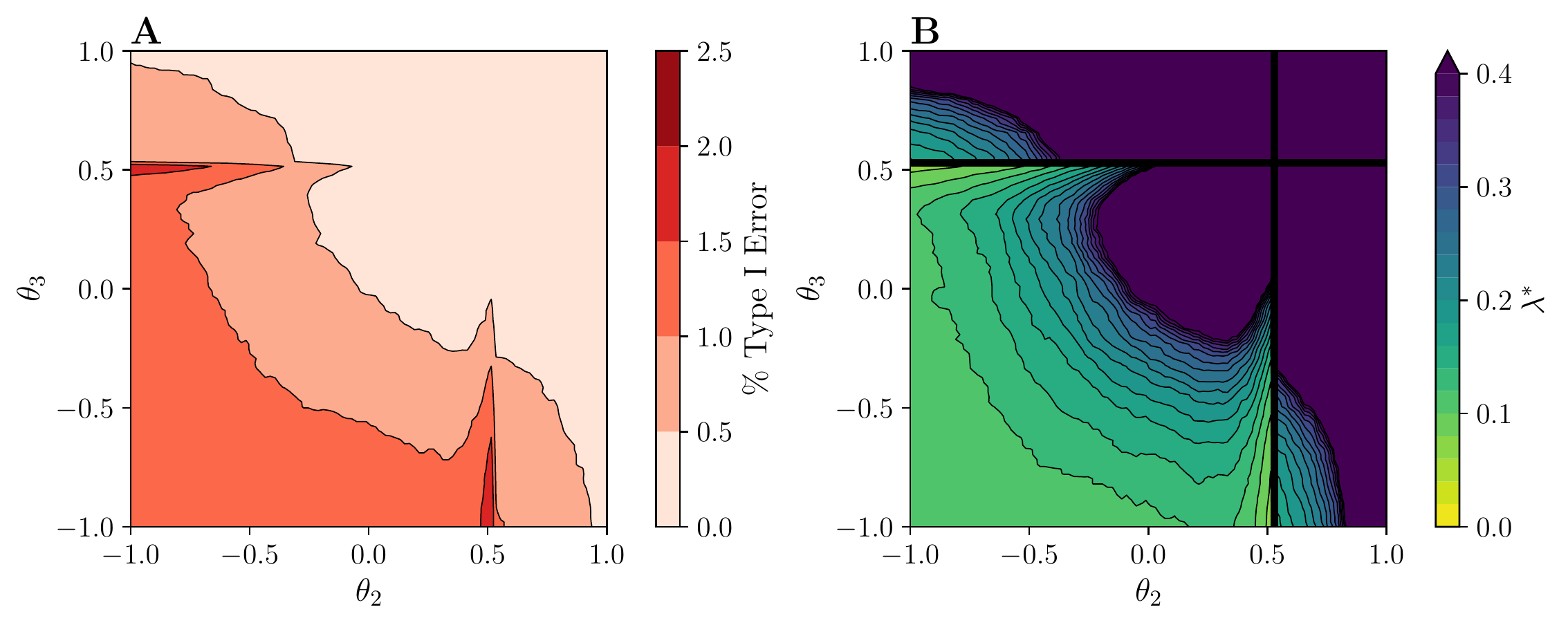}
    \caption{Calibration results of the Phase II/III design. Both plots slice the domain by fixing 2 parameters, $\theta_0 = \theta_1 = 0.533$. Figure \textbf{A} shows the Tilt-Bound (Type I Error upper bound) profile for the selected threshold $\lambda = 0.06253$. Figure \textbf{B} shows the critical value $\lambda^*$ separately for each tile such that its Tilt-Bound is 2.5\%.}
    \label{lewisoutput}
\end{figure}

\begin{figure}
    \centering
    \includegraphics[width=10cm]{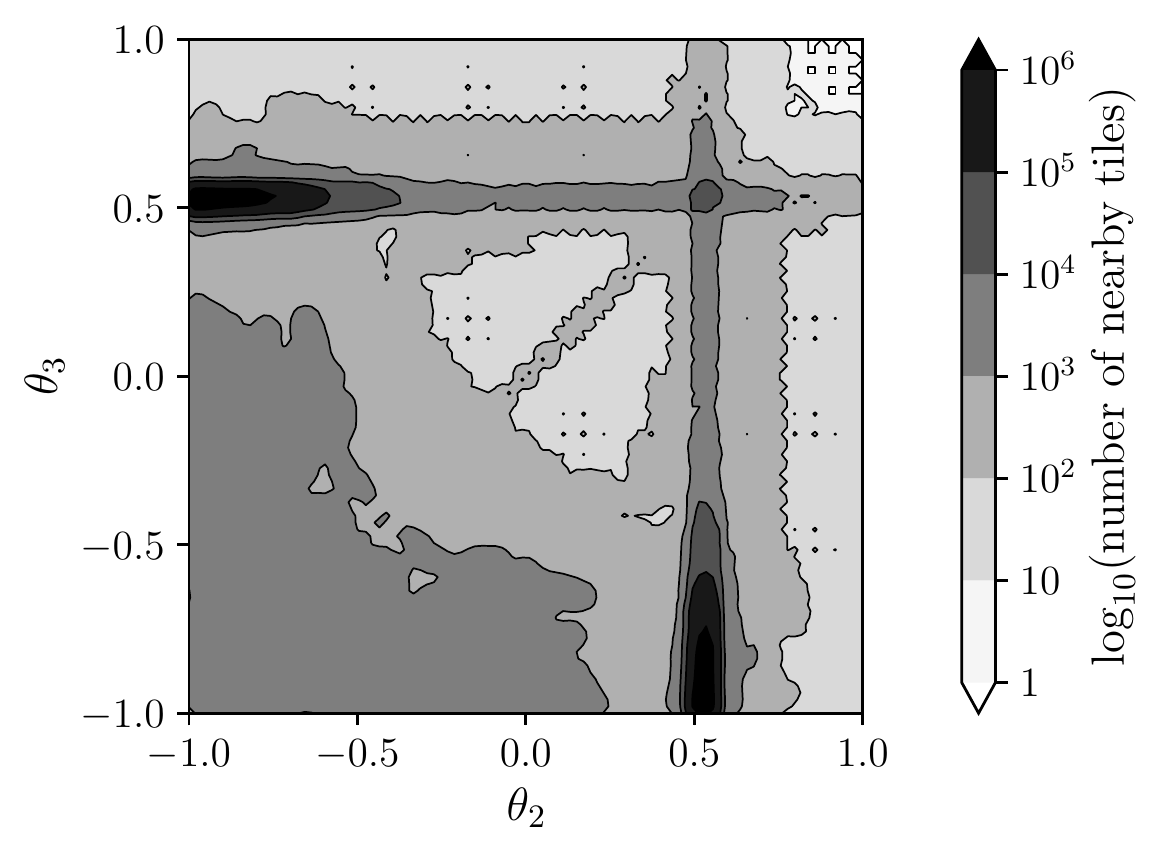}
    \caption{The number of tiles within a ball of radius 0.04 in the $\theta$ space. This shows how the adaptive algorithm assigns many more tiles to regions with high Type I Error. We show a slice of the domain where $\theta_0 = \theta_1 = 0.533$. }
    \label{lewisadagrid}
\end{figure}

Calibration is only computationally tractable for this problem if we can use larger tiles for some (uninteresting) regions of space and small tiles for others. We adaptively use pilot simulations to select the geometry of tiles and number of simulations to perform. The smallest tiles have half-width $0.000488$ whereas the largest tiles have half-width $0.0625$. To grid the entire space at the density of the smallest tiles would require 281 trillion tiles. Instead, we used 38.6 million tiles. Tiles in regions of low Type I Error (FWER) use as few as 2048 simulations, while simulations in regions of high Type I Error use up to 524,288 simulations (\Cref{lewisadagrid}).
In total, adaptive simulation reduces the total number of required simulations from $1.5 \times 10^{20}$ to 960 billion for a total computational savings of approximately 160 million times. As noted above, without adaptivity, the solution here would be impossible even with the largest supercomputers available today. Instead, we are able to produce these results in 5 days with a single Nvidia V100 GPU.

\begin{figure}[h]
    \centering
    \includegraphics[width=0.8\textwidth]{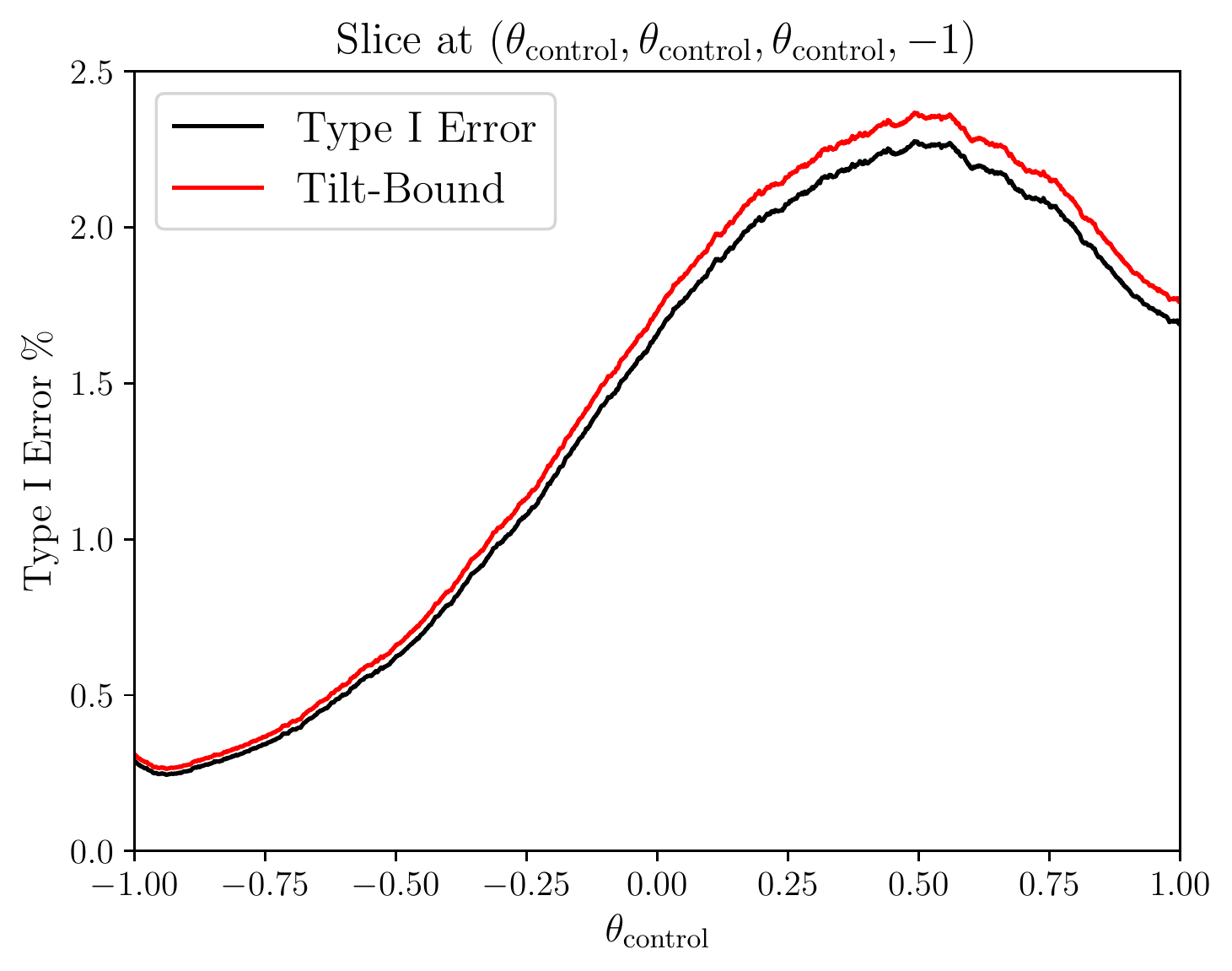}
    \caption{Type I Error estimates, with Tilt-bound from a separate simulation with fixed $\lambda = .06253$, focused on the 1-dimensional set where $\theta_0 = \theta_1 = \theta_2 = \theta_{control}$ and $\theta_3 = -1$. The worst case Type I Error appears when the trio of parameters is set to approximately 0.5.}
    \label{lewis1dcontrolslice}
\end{figure}

We estimate using bootstrapping that at the critical point, the conservative bias due to estimation uncertainty accounts for an expected loss in Type I Error of about $0.16\%$
(out of a total target $2.5\%$); the amount of Type I Error spent on Tilt-Bound CSE from the simulation point to the edges of tiles is $0.06\%$. (We note that the latter quantity is not ``purely slack" because the Type I Error at the simulation point is generally not the worst-case on its tile.)

\begin{figure}[t]
    \centering
    \includegraphics[width=0.8\textwidth]{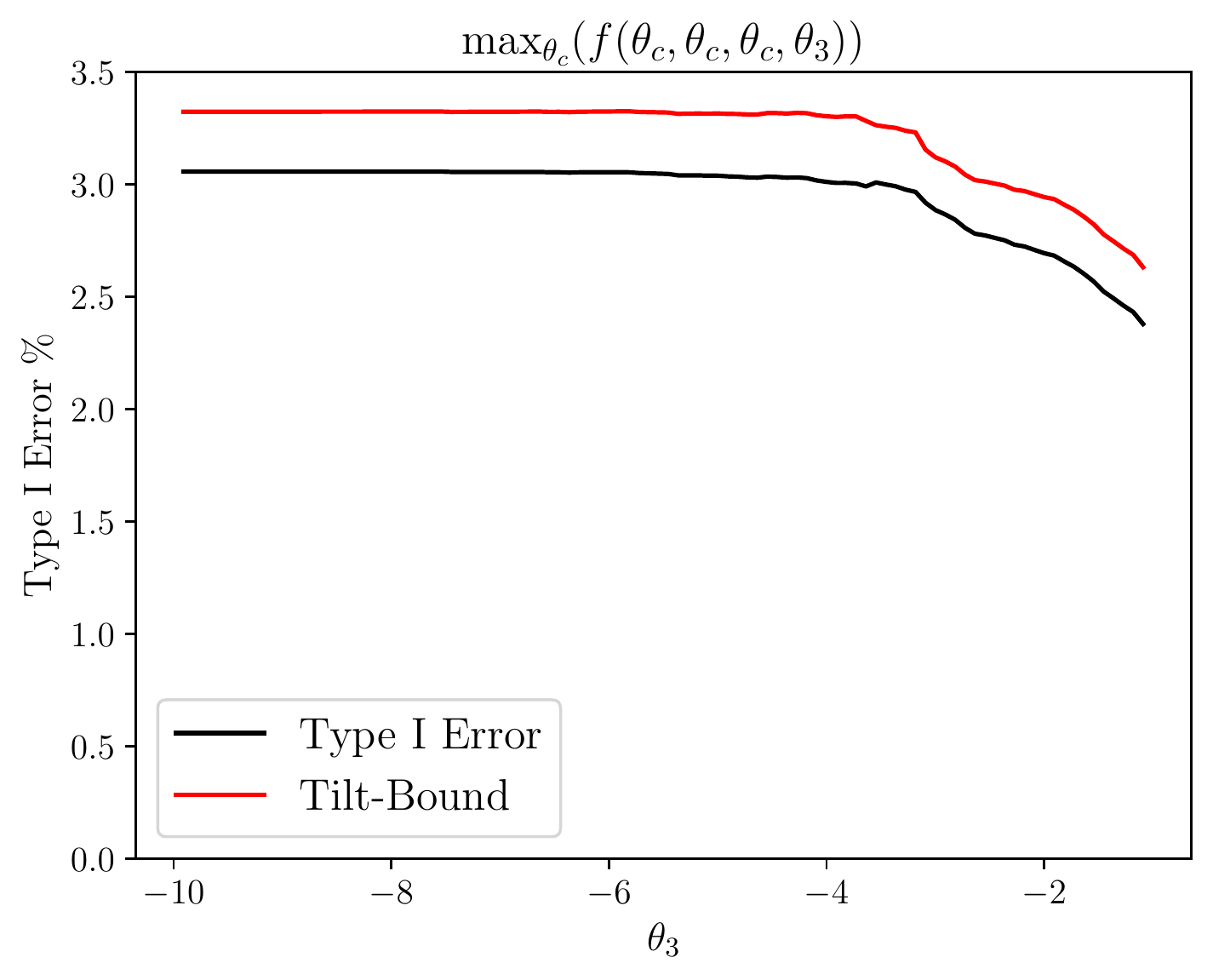}
    \caption{Worst-case Type I Error as a function of $\theta_3$, where the remaining $\theta_i$ are constrained equal to each other and then set to the value which maximizes Type I Error given $\theta_3$. Validation has been performed over this domain with $\lambda$ set to 0.06253; the Tilt-Bound is shown in red.
    In fact, it is possible to extend this graph from the leftmost point $\theta_3 = -10$ $(p_3 = 0.000045)$ all the way to $\theta_3 = -\infty$ $(p_3=0)$ at the cost of a small additional expansion of the bound by about 0.3\% (not shown). This is possible because the sequence of distributions as $\theta_3 \to -\infty$ converges in total variation, and if $\theta_i < -10$, the arm is overwhelmingly likely to dropped before it generates any successes whatsoever. From the left end-point $\theta_3 = -10$, since we can safely assume that this arm would be dropped after observing zero successes out of $n_3 = 50$ patients, the resulting bound would be a penalty of $<0.3\%$ added to the Type I Error at the leftmost point of the figure.}
    \label{lewis1dorthogonal}
\end{figure}


The Tilt-Bound profile of the calibrated design is presented in~\Cref{lewisoutput}. The maximum of the Tilt-Bound occurs at the tile centered at
$\theta = (0.4925, 0.4925, 0.4925, -1.0)$. Surprisingly, the worst Type I Error point \textit{does not} occur at the global null (where all treatments perform equally to control), but rather when one treatment performs poorly.
It appears this ``rogue arm" paradoxically \textit{increases} the chance that other treatments are incorrectly rejected, because the observed heterogeneity causes a reduction in Bayesian borrowing. Unlike the design studied in~\Cref{ssec:berry}, borrowing effects from highly successful arms do not cause Type I Error inflation. The Type I Error profile is non-monotonic with respect to $\theta_i$ and maximized away from the global null.

We remark that with the standard method of simply performing simulations
at each grid point (i.e. without CSE), 
it would be difficult to establish worst-case control.
Even after correctly guessing at the $(\theta, \theta, \theta, -1)$ form of the worst-case point, 
one would still need a 1-dimensional search of Type I Error over $\theta$ (such as~\Cref{lewis1dcontrolslice}).
In contrast, the ``proof-by-simulation" approach establishes Type I Error control over the region of interest unambiguously.

However, CSE guarantees do not necessarily extend \textit{outside} of the region of focus. If $\theta_3$ is permitted to drop below $-1$, as shown in~\Cref{lewis1dorthogonal} the worst-case Type I Error rises above 2.5\% and asymptotes at approximately 3.1\%. If a 2.5\% guarantee were desired for this area, then the initial domain $\Theta$ should have been extended to a wider region at the outset of calibration (although at an additional computational cost). Interestingly, it is technically possible to give a bound which extends~\Cref{lewis1dorthogonal} to the left all the way to $\theta_3 = -\infty$ $(p_3=0)$ at the cost of a small additional penalty. See description of~\Cref{lewis1dorthogonal}.

\section{Discussion}

Proof-by-simulation is a powerful and robust methodology. It places minimal constraints on the structure of the test and sampling decisions, and therefore can successfully validate procedures that defy classical analysis. Continuous Simulation Extension (CSE) enhances simulation-gridding with guarantees, providing an objective standard and removing doubts about potentially missing areas of the null hypothesis space or selective presentation of results.

A practical advantage of CSE is that it analyzes the design as represented in code. Thus, CSE offers designers a ``free pass" to use approximate estimators and inference, and defuses theoretical uncertainties such as convergence of MCMC samplers. In fact, even if the inference code were to contain an undiscovered error, CSE guarantees would remain technically valid. 

To ensure CSE's conservatism is kept small, the number of simulations required grows with the dimension of the parameter space. Our examples demonstrate that up to 4 unknown parameters can be handled using trillions of simulations;  problems with fewer unknown parameters will be easier and require fewer simulations. We anticipate that large-scale computing will only become easier with time. In addition, our framework can be made more efficient with improvements to tile geometry, better adaptive gridding, and extensions to importance sampling. Thus, the set of compatible applications will grow. 

In real medical experiments with adaptive elements, recruitment and logistical uncertainties may leave sample sizes and other ``design parameters" undetermined at the start, to be revealed as the trial goes on; or the trial may undergo significant unplanned modifications. In this case, a specialized form of CSE can ensure maintenance of statistical control by recalibrating ``group-sequentially" at each interim, using the conditional Type I Error principle similar to \cite{koenig2008adaptive} and alternately conditioning on ancillary parameters and re-evaluating conditional Type I Error budget at each stopping-point. However this process may involve many repeated simulations over which simulation error can accumulate; we leave a full elucidation to future work. 

For regulatory applications, a further subjective issue remains: how to justify the modeling family over which CSE will provide its guarantees? The history of previously accepted models is important here, but outside the scope of this paper. 
Yet it bears mentioning that CSE should be easier for regulators to accept than parametric \textit{regression} methods.
To perform inference on a parametrized regression model, such as a confidence interval on the parameters of a logistic regression, requires \textit{estimability} of the model as well as the \textit{relevance} of the model class for inference. CSE relies only on the relevance; therefore even in cases where a parametric regression model is too uncertain or unstable to use, CSE can still be justified. 

We also remark that even if distributional assumptions are unclear, additional evidence of robustness can be given by performing CSE for each of them, or by embedding the models in a larger family. In particular, if a Brownian motion limit applies for the statistics of interest, even simulations taken under the wrong parametric family may offer partial evidence of statistical control. 
For example: consider applying CSE using an incorrect assumption  that data follow $\Gamma(\alpha,\beta)$ outcomes. 
The test statistics for these simulations may follow an approximate $\Normal(\mu,\sigma^2)$ distribution. Then, using CSE to cover a region of $(\alpha,\beta)$ is approximately covering a region of $(\mu,\sigma^2)$ for the test statistic. In this case, using CSE to control Type I Error over a wide range range of $(\alpha,\beta)$ would imply asymptotic control over a wide class of distributions with similar convergence properties.



Creative innovations in design optimization can be supported by CSE tools. For example, CSE can tighten or level out the Type I Error surface of a design. In design optimization, CSE may be useful as an ``inner loop" to ensure that the current version of the design is valid, or to identify and quantify constraint violations. More speculatively, this framework could be used to calibrate ``black-box" design protocols, like using a neural network to determine randomization probabilities or stopping decisions as a function of the sufficient statistics.

But our greatest hope is that these proof-by-simulation techniques can speed up innovation cycles. If properly implemented, we believe proof-by-simulation can reduce the human capital cost of parsing and validating new trial designs, improve regulatory consistency by enabling rigorous satisfaction of objective standards, and offer innovators in design a rapid alternative to peer review for establishing formal claims. 
Yet, much practical work remains to expand our software implementation of CSE for general use. In particular, the case studies in this work and ~\citet{sklar:2022} have been limited to Gaussian and Binomial models. To ensure practical usability, future research should expand these investigations to other model classes, develop introductory materials, continue to improve methodology with techniques such as importance sampling to increase efficiency, and build infrastructure for large-scale simulations and cost estimation to ensure that users find CSE methods cost-effective.
We invite statisticians and regulators to explore our open-source code repository at \cite{ConfirmGithub} and consider its use for the next generation of statistical methodology.

\section{Acknowledgements}

We thank Alex Constantino and Gary Mulder for significant volunteer contributions to our software. Thanks also to Daniel Kang, Art Owen, Narasimhan Balasubramanian, and various regulators and biostatisticians for help and advice. This work was supported by ACX Grants.

\bibliographystyle{unsrtnat}
\bibliography{bibliography}

\appendix
\section{Proof of Main Results}%
\label{sec:appendix-a}

In this section, we provide proofs
of several main theorems for CSE 
stated in the main text.
We first state a few preliminary results.

\begin{lemma}[Basic Properties]%
\label{lem:basic-properties-dq}

Consider the setting of~\Cref{thm:tilt-bound}.
Fix any $\theta_0 \in \Theta$. 
For every $q \geq 1$ and $v \in \R^d$, define 
\begin{align}
    \sD_q 
    &:=
    \set{v \in \R^d : \psi(\theta_0, v, q) < \infty
    }
    \label{eq:Dq:def}
    \\
    \varphi_q(v)
    &:=
    \frac{\psi(\theta_0, v, q)}{q}
    -
    \psi(\theta_0, v, 1)
    \label{eq:varphiq:def}
    \\
    \sI_v 
    &:= 
    \set{q \geq 1 : \psi(\theta_0, v, q) < \infty}
    \label{eq:Iv:def}
\end{align}
Let $q_1$, $q_2$ be any pair such that $1 \leq q_1 < q_2$. 
Then, the following statements hold:
\begin{enumerate}[label=(\roman*)]
\item\label{lem:prop:set-increase} $\mathcal{D}_{q_1} \supseteq \mathcal{D}_{q_2}$.
\item\label{lem:prop:convex} If $v\mapsto \Delta_{\theta_0}(v, x)$ is linear,
then $\mathcal{D}_{q}$ is convex.
\item\label{lem:prop:nonneg} $\mathcal{D}_q = \{v: 0 \leq \varphi_q(v) < \infty\}$.
\item\label{lem:prop:convex2} $\mathcal{I}_v$ is convex and 
$q \mapsto \psi(\theta_0, v, q)$ is convex on $\mathcal{I}_v$.
It is strictly convex if $\Delta_{\theta_0}(v, X)$ is not constant
$P_{\theta_0}$-a.s.
\end{enumerate}

\begin{proof}

If $v \in \mathcal{D}_{q_2}$, then
$\psi(\theta_0, v, q_2) < \infty$.
By Jensen's Inequality,

\begin{align}
    \norm{e^{\Delta_{\theta_0}(v,X)}}_{L^{q_1}(P_{\theta_0})}
    &\leq
    \norm{e^{\Delta_{\theta_0}(v,X)}}_{L^{q_2}(P_{\theta_0})}
    \nonumber
    \\\implies
    \frac{1}{q_2} \psi(\theta_0, v, q_1)
    &\leq
    \frac{1}{q_2} \psi(\theta_0, v, q_2)
    < \infty
\label{eq:domain_contain}
\end{align}
Thus, we have~\labelcref{lem:prop:set-increase}.

Without loss of generality, now assume that $\mathcal{D}_q$ is non-empty.
For any $v_1, v_2$ and any $\lambda \in (0,1)$,
\begin{align*}
    \EEE_{\theta_0}\br{
    e^{q \Delta_{\theta_0}(\lambda v_1 + (1-\lambda) v_2, X)}
    }
    &\leq
    \EEE_{\theta_0}\br{
    e^{q \Delta_{\theta_0}(v_1, X)}
    }^{\lambda}
    \EEE_{\theta_0}\br{
    e^{q \Delta_{\theta_0}(v_2, X)}
    }^{1-\lambda}
\end{align*}
by H\"{o}lder's Inequality and linearity of $\Delta_{\theta_0}(\cdot, X)$.
Taking $\log$ on both sides,
\begin{align*}
    \psi(\theta_0, \lambda v_1 + (1-\lambda)v_2, q)
    &\leq
    \lambda \psi(\theta_0, v_1, q)
    +
    (1-\lambda) \psi(\theta_0, v_2, q)
    < \infty
\end{align*}
Thus, $\lambda v_1 + (1-\lambda) v_2 \in \mathcal{D}_q$.
This proves~\labelcref{lem:prop:convex}.

Next, consider $\varphi_q$.
Note that $\psi(\theta_0, v, 1)$ cannot be $-\infty$ because $P_{\theta_0 + v}$ is a probability measure absolutely continuous to $P_{\theta_0}$. Also, $\mathcal{D}_{1} \supseteq \mathcal{D}_{q}$ by~\labelcref{lem:prop:set-increase}.
Therefore, for $v \in \mathcal{D}_{q}$, 
we must have $\varphi_1(v) = 0$. 
Further,~\labelcref{eq:domain_contain} shows that $q \mapsto \frac{\psi(\theta_0, v, q)}{q}$ is non-decreasing. 
Thus, 
\begin{align*}
    \mathcal{D}_q 
    := 
    \set{v: \abs{\varphi_q(v)} < \infty} 
    = 
    \set{v: 0 \leq \varphi_q(v) < \infty} 
\end{align*}
Thus, this proves~\labelcref{lem:prop:nonneg}.

Finally, we show that $\mathcal{I}_v$ is convex.
For notational ease, let $Z := \Delta_{\theta_0}(v, X)$.
For any $q_1 \neq q_2 \in \mathcal{I}_v$ and $\lambda \in (0, 1)$,
we apply H\"{o}lder's Inequality to get that
\begin{align*}
    \EEE_{\theta_0}\br{
        e^{\pr{\lambda q_1 + (1-\lambda) q_2} Z}
    }
    &\leq
    \EEE_{\theta_0}\br{
        e^{q_1 Z}
    }^{\lambda}
    \EEE_{\theta_0}\br{
        e^{q_2 Z}
    }^{1-\lambda}
\end{align*}
Taking $\log$ on both sides,
\begin{align*}
    \psi(\theta_0, v, \lambda q_1 + (1-\lambda) q_2)
    &\leq
    \lambda \psi(\theta_0, v, q_1)
    + (1-\lambda) \psi(\theta_0, v, q_2)
    < \infty
\end{align*}
Hence, $\lambda q_1 + (1-\lambda) q_2 \in \mathcal{I}_v$.
This proves that $\mathcal{I}_v$ is convex 
and $q\mapsto \psi(\theta_0, v, q)$ is convex on $\mathcal{I}_v$.
Note that H\"{o}lder's Inequality is an equality
if and only if there exists $\alpha, \beta \geq 0$
not both zero such that
\begin{align*}
    \alpha e^{q_1 Z} 
    =
    \beta e^{q_2 Z}
\end{align*}
This can only occur if $q_1 = q_2$ or $Z$ is constant
$P_{\theta_0}$-a.s.
Hence, if $Z$ is additionally not constant $P_{\theta_0}$-a.s.,
then $q \mapsto \psi(\theta_0, v, q)$
is strictly convex on $\mathcal{I}_v$.
This proves~\labelcref{lem:prop:convex2}.
\end{proof}
\end{lemma}

\begin{lemma}[Monotonicity of $\tilde{\varphi}_q$]%
\label{lem:gq-is-monotonic}
Consider the setting of~\Cref{thm:tilt-bound-qcv}.
Fix any $\theta_0 \in \Theta\subseteq \R^d$, 
$v_0 \in \R^d$,
$u \in \R^d$ such that $u \perp v_0$, 
and $q \geq 1$.
Define
\begin{align*}
    \sD 
    &:= 
    \set{t \in \R : v_0+tu \in \sD_q}
    \\
    \tilde{\varphi}_q(h)
    &:=
    \varphi_q(v_0 + hu)
\end{align*}
where $\sD_q$ and $\varphi_q$ are 
given by~\labelcref{eq:Dq:def,eq:varphiq:def}, respectively.
Then, for all $h \in \mathcal{D} \cap (0,\infty)$,
$\frac{\partial}{\partial h} \tilde{\varphi}_q(h) \geq 0$. 
\begin{proof}
Without loss of generality, assume $\sD \cap (0,\infty)$ is non-empty and consider any $h \in \sD \cap (0,\infty)$.
We begin with the expression from the chain rule:
\begin{align*}
    \frac{\partial}{\partial h}
    \tilde{\varphi}_q(h)
    &=
    \nabla \varphi_q(v_0 + hu)^\top u
    \\
    \nabla \varphi_q(v)
    &=
    \frac{\nabla_v \psi(\theta_0, v, q)}{q}
    -
    \nabla_v \psi(\theta_0, v, 1)
\end{align*}
Since $h > 0$, letting $u_h := hu$,
we may equivalently show that
\begin{align*}
    h \frac{\partial}{\partial h} \tilde{\varphi}_q(h)
    =
    \nabla \varphi_q(v_0 + u_h)^\top u_h \geq 0
\end{align*}
To establish the above, 
by an integration argument, it suffices to show that
\begin{align*}
    \frac{\partial}{\partial t} 
    \pr{
        \frac{\nabla_v \psi(\theta_0, v_0 + u_h, t)^\top u_h}{t}
    }
    \geq 0
\end{align*}
for $t \in [1, q]$.
Note that all terms are well-defined for any $t \in [1,q]$
since 
\[
v_0 + u_h \in \mathcal{D}_{q} = \bigcap\limits_{t \in [1,q]} \mathcal{D}_t
\]
by~\Cref{lem:basic-properties-dq}~\labelcref{lem:prop:set-increase}.
By integrating, we then have that
\begin{align*}
    h \frac{\partial}{\partial h}
    \tilde{\varphi}_q(h)
    &=
    \int_1^q 
    \frac{\partial}{\partial t} 
    \pr{
        \frac{\nabla_v \psi(\theta_0, v_0+u_h, t)^\top u_h}{t}
    }
    dt
    \geq
    0
\end{align*}

For any $v \in \R^d$, 
define a family of distributions 
\[
\mathcal{Q}_v := \set{Q_{t, v} : t \geq 1,\, \psi(\theta_0, v, t) < \infty}
\]
where $Q_{t,v}$ is given by
\begin{align*}
    dQ_{t, v}(x)
    &:=
    \exp\br{
    t \Delta(v, x) - \psi(\theta_0, v, t)
    }
    dP_{\theta_0}(x)
\end{align*}
Using the definition of $\psi$ and $Q_{t,v}$, we have that
\begin{align*}
    \nabla_v \psi(\theta_0, v, t)
    &=
    t e^{-\psi(\theta_0, v, t)}
    \EEE_{\theta_0}\br{
    \nabla_v \Delta(v, X) e^{t \Delta(v, X)}
    }
    \\&=
    t \EEE_{Q_{t,v}} \br{
        \nabla_v \Delta(v, X)
    }
\end{align*}
Hence, for any $t \in [1,q]$,
\begin{align}
    \frac{\nabla_v \psi(\theta_0, v_0+u_h, t)^\top u_h}{t}
    &=
    \EEE_{Q_{t, v_0+u_h}}\br{
    \nabla_v \Delta(v_0+u_h, X)^\top u_h
    }
    \label{eq:tilt-bound-max-vertex:qexpr}
\end{align}

Note that $Q_{t,v}$ forms an exponential family
with natural parameter $t$,
sufficient statistic $\Delta(v, x)$,
and log-partition function $\psi(\theta_0, v, t)$
with base measure $P_{\theta_0}$.
By standard results for exponential family,
the derivative of~\labelcref{eq:tilt-bound-max-vertex:qexpr}
with respect to $t$ is
\begin{align*}
    \frac{\partial}{\partial t}
    \pr{\frac{\nabla_v \psi(\theta_0, v_0 + u_h, t)^\top u_h}{t}}
    &=
    \cov_{Q_{t, v_0+u_h}}\pr{
        \nabla_v \Delta(v_0+u_h, X)^\top u_h,\;
        \Delta(v_0+u_h, X)
    }
    \\&=
    u_h^\top 
    \left[\var_{Q_{t, v_0+u_h}} W(X)\right]
    (v_0 + u_h)
\end{align*}
Let $P_0$ denote the projection matrix onto 
$\text{span}(v_0)^\perp$.
Then, since $v_0 \perp u_h$ by hypothesis,
we have $P_0 (v_0 + u_h) = P_0 u_h = u_h$.
Consequently,
\begin{align}
    u_h^\top \left[ \var_{Q_{t, v_0+u_h}} W(X) \right] \, (v_0 + u_h)
    =
    (v_0+u_h)^\top P_0 \left[ \var_{Q_{t, v_0+u_h}} W(X)\right] \, (v_0+u_h)
    \label{eq:tilt-bound-max-vertex-cond}
\end{align}
Since $P_0$ and $\var_{Q_{t, v_0+u_h}}W(X)$ are
both positive semi-definite, 
their product must have all non-negative eigenvalues.
Hence, the right-side of~\labelcref{eq:tilt-bound-max-vertex-cond}
is non-negative.
This proves that $\frac{\partial}{\partial h} \tilde{\varphi}_q(h) \geq 0$ on $\mathcal{D} \cap (0, \infty)$.

\end{proof}
\end{lemma}

\begin{lemma}\label{lem:strict-quasiconvex}
Let $f(x) = \frac{f_1(x)}{f_2(x)}$ where
$f_1 : C \mapsto \R$ and $f_2: C \mapsto (0, \infty)$
for some convex set $C$ of a vector space.
If $f_1$ is strictly convex and $f_2$ concave,
then $f$ is strictly quasi-convex.

\begin{proof}
Fix any $x \neq y \in C$ and $\lambda \in (0,1)$.
Then, 
\begin{align*}
    f(\lambda x + (1-\lambda) y)
    &<
    \frac{\lambda f_1(x) + (1-\lambda) f_1(y)}{\lambda f_2(x) + (1-\lambda) f_2(y)}
    \\&=
    \alpha f(x) + (1-\alpha) f(y)
    \leq
    \max(f(x), f(y))
\end{align*}
where $\alpha = \frac{\lambda f_2(x)}{\lambda f_2(x) + (1-\lambda) f_2(y)}$.
\end{proof}
\end{lemma}

\subsection{Proof of~\Cref{thm:tilt-bound-qcp}}
\label{appendix:thm:tilt-bound-qcp}
\begin{proof}
Fix $\theta_0 \in \Theta$, 
a set $S \subseteq \R^d$, 
and $a \geq 0$. 
Without loss of generality, assume $a > 0$.
If $a = 0$, we may simply set $q^* \equiv \infty$.
Note that any $q \geq 1$ achieves the minimum,
so the minimizer is not unique.

Let $\mathcal{I}_v$ be as in~\labelcref{eq:Iv:def}.
Define $\mathcal{I} := \bigcap\limits_{v \in S} \mathcal{I}_v$.
By~\Cref{lem:basic-properties-dq}~\labelcref{lem:prop:convex2}, 
$\sI_v$ is convex for every $v \in S$ so 
$\sI$ is convex as well.
It suffices to minimize
\[q \mapsto \sup\limits_{v\in S} U(\theta_0, v, q, a)\]
on $\mathcal{I}$ since it is infinite on $\mathcal{I}^c$.
Without loss of generality,
assume that $\mathcal{I}$ is non-empty,
or equivalently,
$q \mapsto \sup\limits_{v\in S} U(\theta_0, v, q, a)$
is not identically infinite.
Otherwise, we may set $q^* \equiv \infty$
and the minimizer is not unique.

We show that 
$q \mapsto \sup\limits_{v \in S} U(\theta_0, v, q, a)$ is quasi-convex on $\mathcal{I}$
and is strict if $S$ is finite.
Note that 
\begin{align*}
    \sup\limits_{v \in S} U(\theta_0, v, q, a)
    &=
    a \cdot
    \exp\br{
    \sup\limits_{v \in S} \br{
    \frac{\psi(\theta_0, v, q) - \log(a)}{q}
    - \psi(\theta_0, v, 1) 
    }
    }
\end{align*}
To establish the desired claim, 
we show that $\tilde{\psi}(q)$
is (strictly, if finite $S$) quasi-convex on $\mathcal{I}$ where
\begin{align*}
    \tilde{\psi}(q)
    &:=
    \sup\limits_{v \in S} 
    \tilde{\psi}_v(q)
    \\
    \tilde{\psi}_v(q)
    &:=
    \frac{\psi(\theta_0, v, q) - \log(a)}{q}
    - \psi(\theta_0, v, 1)
\end{align*}
It suffices to show that
$q \mapsto \tilde{\psi}_v(q)$ is strictly
quasi-convex on $\mathcal{I}_v$ for every $v \in S$.
Note that $\tilde{\psi}_v(q) = \frac{f_1(q)}{f_2(q)}$
where 
\begin{align*}
    f_1(q) 
    &:= 
    \psi(\theta_0, v, q) - \log(a) 
    - q \psi(\theta_0, v, 1)
    \\
    f_2(q)
    &:=
    q
\end{align*}
\Cref{lem:basic-properties-dq} 
shows that $q \mapsto \psi(\theta_0, v, q)$ 
is strictly convex on $\mathcal{I}_v$,
which directly shows the same for $f_1(q)$.
Then,~\Cref{lem:strict-quasiconvex}
shows that $\tilde{\psi}_v(q)$ is strictly quasi-convex.

Finally, since $\tilde{\psi}$ is (strictly, if finite $S$)
quasi-convex on $\mathcal{I}$,
we have the following quasi-convex program:
\begin{align*}
    &\minimize\limits_q \tilde{\psi}(q) \\
    &\subjto \quad q \in \mathcal{I}
\end{align*}
This proves the existence (and uniqueness for finite $S$)
of a global minimizer $q^* \in \mathcal{I} \subseteq [1, \infty]$.
\end{proof}

\subsection{Proof of~\Cref{thm:tilt-bound-qcv}}\label{appendix:thm:tilt-bound-qcv}
\begin{proof}
Fix any $\theta_0 \in \Theta$. 
For every $q \geq 1$, define 
$\sD_q$ as in~\labelcref{eq:Dq:def}.
By~\Cref{lem:basic-properties-dq}~\labelcref{lem:prop:convex,lem:prop:nonneg}, 
$\sD_{q} \equiv \set{v : \abs{\varphi_q(v)} < \infty}$ 
is convex,
where $\varphi_q$ is as in~\labelcref{eq:varphiq:def}. 
Now, fix any $q \geq 1$.
To establish quasi-convexity of the Tilt-Bound~\labelcref{eq:tilt-bound:def} as a function of $v$, it therefore suffices to show that $\varphi_q(v)$ is quasi-convex along all one-dimensional line segments contained in $\mathcal{D}_{q}$. Let $a$, $b$ be any two points in $\mathcal{D}_{q}$, and $\overline{ab}$ the line segment joining them. We may consider the directional unit vector
$u = \frac{b-a}{\|b-a\|}$,  extend the line segment to make a full line $\overleftrightarrow{ab}$, and then drop a perpendicular vector $v_0$ 
which measures the distance from the origin to $\overleftrightarrow{ab}$. 
Hence, without loss of generality, 
it suffices to show that
$h \mapsto \varphi_q(v_0 + hu)$ is quasi-convex 
for all vectors $v_0$ and 
$u$ such that $u \perp v_0$ and 
$h \in \mathcal{D}_{q, v_0, u} := 
\set{t \in \R : v_0 + tu \in \mathcal{D}_q}$. 
Note that $\mathcal{D}_{q, v_0, u}$ is an interval 
by convexity of $\mathcal{D}_{q}$.

Without loss of generality, 
fix any such $v_0$ and $u$.
Denote $\mathcal{D} \equiv \mathcal{D}_{q, v_0, u}$ for 
notational ease.
Define $\tilde{\varphi}_q(h) := \varphi_q(v_0 + hu)$. 
We wish to show that $\tilde{\varphi}_q$ 
is quasi-convex on $\mathcal{D}$.
\Cref{lem:gq-is-monotonic} shows that
$\frac{\partial}{\partial h} \tilde{\varphi}_q(h) \geq 0$ 
for all $h \in \mathcal{D} \cap (0, \infty)$,
which shows that $\tilde{\varphi}_q(h)$ is non-decreasing
for $h\in \mathcal{D} \cap (0, \infty)$.
Note that since $\Delta(\cdot, X)$ is linear,
we have by standard exponential family that
for every $\tilde{q} \geq 1$,
$v \mapsto \psi(\theta_0, v, \tilde{q})$ is continuous on $\sD_{\tilde{q}}$.
Then, by~\Cref{lem:basic-properties-dq}~\labelcref{lem:prop:set-increase}, 
$v \mapsto \psi(\theta_0, v, 1)$ is continuous on $\sD_q$ so
$\varphi_q$ is continuous on $\mathcal{D}_q$.
Hence, $\tilde{\varphi}_q$ is non-decreasing on
$\mathcal{D} \cap [0,\infty)$.
Therefore, applying the same logic for $-u$,
$\tilde{\varphi}_q$ must either 
be monotone on $\mathcal{D}$
or decreasing from the left end-point of $\mathcal{D}$
to $0$ then increasing to the right end-point of $\mathcal{D}$.
Thus, $\tilde{\varphi}_q$ is quasi-convex 
along the segment $\mathcal{D}$, as needed.
\end{proof}

\subsection{Proof of~\Cref{thm:calibration}}
\label{appendix:thm:calibration}

\begin{proof}
Define $\tilde{U}_i := f(S_i)$ for every $i=1,\ldots, N$.
Then, by monotonicity of $f$,
\begin{align}
    \EEE\br{f(S_{(\floor{(N+1) \alpha})})}
    &=
    \EEE\br{\tilde{U}_{(\floor{(N+1) \alpha})}}
    \label{eq:calibration:sub-unif}
\end{align}

We claim that $\tilde{U}_i$ are sub-uniform, that is,
they satisfy the following property:
\begin{align*}
    \prob\pr{\tilde{U} \leq x} \geq x
\end{align*}
for every $x \in [0,1]$.
To prove this, we first define a pseudo-inverse function
\begin{align*}
    f^{-1}(y) 
    :=
    \inf\set{\lambda : f(\lambda) \geq y}
\end{align*}
for all $y\in [0,1]$.
Note that 
\begin{align*}
    f(f^{-1}(y)) 
    \leq
    y
    \leq
    f_+(f^{-1}(y))
\end{align*}
by left-continuity of $f(\lambda)$.
Hence, for any $x \in [0,1]$,
\begin{align*}
    \prob\pr{\tilde{U} \leq x}
    &=
    \prob\pr{f(S) \leq x}
    \geq
    \prob\pr{S \leq f^{-1}(x)}
    =
    f_+(f^{-1}(x))
    \geq 
    x
\end{align*}

By considering a possibly enlarged probability space,
we may construct a uniform random variable $U_i \sim U(0,1)$
for each $\tilde{U}_i$ such that $\tilde{U}_i \leq U_i$.
In particular, we have that $\tilde{U}_{(k)} \leq U_{(k)}$ for any $k=1,\ldots, N$.
Continuing from~\labelcref{eq:calibration:sub-unif},
\begin{align*}
    \EEE\br{\tilde{U}_{(\floor{(N+1)\alpha})}}
    &\leq
    \EEE\br{U_{(\floor{(N+1)\alpha})}}
\end{align*}
Note that $U_{(k)} \sim \Beta\pr{k, N-k+1}$ for every $k=1,\ldots, N$.
Hence,
\begin{align*}
    \EEE\br{U_{(\floor{(N+1)\alpha})}}
    &=
    \frac{\floor{(N+1)\alpha}}{N+1}
\end{align*}

Next, note that
\begin{align*}
    f(s) &= f_+(s) - \Delta f(s) \\
    \implies
    \EEE\br{f(S_{(\floor{(N+1)\alpha})})}
    &=
    \EEE\br{f_+(S_{(\floor{(N+1)\alpha})})}
    -
    \EEE\br{\Delta f(S_{(\floor{(N+1)\alpha})})}
\end{align*}
Hence, it suffices to show that
\begin{align*}
    \EEE\br{f_+(S_{(\floor{(N+1)\alpha})})}
    &\geq
    \frac{\floor{(N+1)\alpha}}{N+1}
\end{align*}
By a similar argument as before,
we construct a pseudo-inverse
\begin{align*}
    f_+^{-1}(y)
    :=
    \inf\set{\lambda : f_+(\lambda) > y}
\end{align*}
so that
\begin{align*}
    f(f_+^{-1}(y))
    &\leq
    y
    \leq
    f_+(f_+^{-1}(y))
\end{align*}
Hence, for any $x \in [0,1)$ and $x < y \leq 1$,
\begin{align*}
    \prob\pr{f_+(S) < y}
    \leq
    \prob\pr{S < f_+^{-1}(y)}
    =
    f(f_+^{-1}(y))
    \leq
    y
\end{align*}
Taking $y \downarrow x$,
\begin{align*}
    \prob\pr{f_+(S) \leq x}
    \leq
    x
\end{align*}
Since $f_+ \in [0,1]$, the result trivially holds for $x=1$.
Hence, $f_+(S)$ is super-uniform.
This implies that
\begin{align*}
    \EEE\br{f_+(S_{(\floor{(N+1)\alpha})})}
    &\geq
    \EEE\br{U_{(\floor{(N+1)\alpha})}}
    =
    \frac{\floor{(N+1)\alpha}}{N+1}
\end{align*}
This concludes the proof of~\labelcref{eq:calibration:tight-mean}.
     
We now prove~\labelcref{eq:calibration:tight-var}.
For notational ease, let $k := \floor{(N+1)\alpha}$.
Then,
\begin{align*}
    \var f(S_{(k)})
    &=
    \EEE\br{f(S_{(k)})^2}
    -
    \EEE\br{f(S_{(k)})}^2
    \\&\leq
    \EEE\br{U_{(k)}^2} 
    -
    \EEE\br{f(S_{(k)})}^2
    \\&\leq
    \var{U_{(k)}}
    +
    \pr{\frac{k}{N+1}}^2
    -
    \pr{\frac{k}{N+1} - \EEE\br{\Delta f(S_{(k)})}}^2
    \\&=
    \var{U_{(k)}}
    +
    \EEE\br{\Delta f(S_{(k)})}
    \pr{\frac{2k}{N+1} - \EEE\br{\Delta f(S_{(k)})}}
\end{align*}
Since $\var U_{(\floor{(N+1)\alpha})} = O(\frac{1}{N})$,
we have the desired claim.
\end{proof}
\section{Phase II/III Design Details}
\label{sec:appendix-d}

The Phase II component begins with 4 arms, 
including 1 arm for control $(a_0)$ and 3 treatment arms $(a_1, a_2, a_3)$, and randomizes a maximum of 400 patients. 
Outcomes are assumed to be immediately observable, 
and distributed $\Bern(p_i)$ for each arm $a_i$. 
Randomization in the Phase II portion is blocked
$1:1:1:1$ for the first 200 patients. An interim analysis occurs after the first 200 patients and then again after every 100 subsequent patients. In the interim analysis each remaining non-control arm is analyzed according to two metrics, $p^i_{\text{best}}$ and $p^i_{\text{success}}$, according to the Bayesian hierarchical model presented in~\Cref{ssec:berry}. The metric $p^i_{\text{success}}$ is meant to approximate the conditional power if arm $a_i$ were accelerated to Phase III. 
More precisely, $p^i_{\text{success}}$ equals the probability under the current Bayesian posterior model that if 200 patients are added to both of the arms $a_i$ and $a_0$, i.e. adding a completed Phase III dataset to the analysis, the resulting final posterior will have 
$\prob\pr{p_i > p_0} > 95\%$. 
This probability is pre-computed by simulation for a grid of possible data values, so that during calibration a rapid approximation can be made by interpolating a lookup table. 
If $p^i_{\text{success}}> 70\%$ 
for any treatment, Phase II concludes with a selection and the current best arm is accelerated to the Phase III portion. Otherwise, there is an assessment of which, if any, arms should be dropped. For $i=1,2,3$, if $p^i_{\text{best}} := \prob\pr{p_i = \max \limits_j p_j} < 15\%$, then arm $a_i$ is dropped for futility and future patients in the Phase II will be split evenly among the remaining arms with fractional patients thrown out. If all treatment arms are dropped, the trial ends in futility. If the Phase II portion reaches its third and final analysis ($\approx 400$ patients randomized), the selection threshold for the best arm to progress to Phase III is lowered to $p^i_{\text{success}}> 60\%$. If no arm achieves this, the trial stops for futility.

If an arm is selected for Phase III, then up to 400 patients will be block-randomized 1:1 between the selected treatment and control. When 200 patients have been thus randomized, an interim analysis is performed. If at this interim $\prob\pr{p_i > p_0} > 95\%$ for the selected $i$, then success is declared immediately. 
If $p^i_{\text{success}} < 20\%$, where $p^i_{\text{success}}$ is the current posterior probability that if 100 patients are added to each $a_i$ and $a_0$ the resulting posterior will have 
$\prob\pr{p_i > p_0} > 95\%$, the Phase III stops for futility. 
If neither occurs, we progress to the final analysis with a final block of 200 block-randomized patients. The success criterion at this final analysis will be a threshold $\lambda$ for the final posterior probability of superiority $\prob\pr{p_i > p_0} > 1 - \lambda$. $\lambda$ is left as a tuning parameter, and is used to calibrate the design for Type I Error control. The selected $\lambda$ was $6.253\%$.

\end{document}